\documentclass[10pt,twocolumn,twoside]{IEEEtran}
\usepackage{amsfonts,dsfont,amssymb,amsbsy,amsmath,paralist,theorem,bm,ifthen,color}
\usepackage[pdfstartview=FitH,bookmarksnumbered,unicode,bookmarksopen=true]{hyperref}
\usepackage{graphicx}
\usepackage{epstopdf}
\usepackage{multirow}
\usepackage{subfigure,relsize}
\usepackage{mathrsfs}
\usepackage{bm}
\usepackage{stfloats}
\usepackage{url}
\usepackage[noadjust]{cite}
\usepackage{cases,booktabs}
\usepackage{graphicx,algorithmic,algorithm,relsize}
\usepackage[justification=centering]{caption} 
\usepackage[table]{xcolor}

\newtheorem{lem}{Lemma}

\newtheorem{rem}{Remark}

\newcommand\pin{\ensuremath{{\rm Pin}}}

\renewcommand\sb{\ensuremath{{\bf s}}}

\newcommand\hb{\ensuremath{{\bf h}}}

\newcommand\gb{\ensuremath{{\bf g}}}

\newcommand\vb{\ensuremath{{\bf v}}}

\newcommand\psib{\ensuremath{{\bm \psi}}}

\def\CN{{{\mathcal{CN}}}}

\newcommand\SNR{\ensuremath{{\sf SNR}}}
\newcommand\SINR{\ensuremath{{\sf SINR}}}

\newcommand\Cs{\ensuremath{{\mathbb{C}}}}

\newcommand\Nset  {\ensuremath{{\mathcal{N}}}}

\newcommand\st    {\ensuremath{{\rm subject~to}}}

\graphicspath{{fig/}}

\definecolor{green}{RGB}{34	195	46}
\definecolor{red}{RGB}{220 0 0}

\usepackage{graphicx} 
\allowdisplaybreaks[4]

\title{QoS-Aware NOMA Design for Downlink Pinching-Antenna Systems}

\author{Yanqing Xu, \IEEEmembership{Member, IEEE,}
        Zhiguo Ding, \IEEEmembership{Fellow, IEEE,}
        Donghong Cai, \IEEEmembership{Senior Member, IEEE,}\\
        and Vincent W.S. Wong, \IEEEmembership{Fellow, IEEE}
        \thanks{\smaller[1] The work of Y. Xu  was supported by the NSFC, China under Grant No. 62201486. The work of D. Cai was supported by the Basic and Applied Basic Research Foundation of Guangdong province  under Grant No. 2024A1515012398. {\textit{(Corresponding author: Yanqing Xu.)}}}
        \thanks{\smaller[1] Y. Xu is with the School of Science and Engineering, The Chinese University of Hong Kong, Shenzhen, 518172, China (email: xuyanqing@cuhk.edu.cn).}
        \thanks{\smaller[1] Z. Ding is with the University of Manchester, Manchester, M1 9BB, UK. (email: zhiguo.ding@manchester.ac.uk).} 
        \thanks{\smaller[1] D. Cai is with College of Cyber Security, Jinan University, Guangzhou, China. (email: dhcai@jnu.edu.cn).} 
        \thanks{\smaller[1]  V. W.S. Wong is with the Department of Electrical and Computer Engineering, The University of British Columbia, Vancouver, Canada. (email: vincentw@ece.ubc.ca).} 
        \thanks{\smaller[1] This work has been submitted to the IEEE for possible publication. Copyright may be transferred without notice, after which this version may no longer be accessible.} 
}

\date{\today}

\begin{document}

\maketitle

\begin{abstract}
Pinching antennas, implemented by applying small dielectric particles on a waveguide, have emerged as a promising flexible-antenna technology ideal for next-generation wireless communications systems. Unlike conventional flexible-antenna systems, pinching antennas offer the advantage of creating line-of-sight (LoS) links by enabling antennas to be activated on the waveguide at a position close to the users. This paper investigates a typical two-user non-orthogonal multiple access (NOMA) downlink scenario, where multiple pinching antennas are activated on a single dielectric waveguide to assist NOMA transmission. We formulate the problem of maximizing the data rate of one user subject to the quality-of-service (QoS) requirement of the other user by jointly optimizing the antenna positions and power allocation coefficients. The formulated problem is nonconvex and difficult to solve due to the impact of antenna positions on large-scale path loss and two types of phase shifts, namely in-waveguide phase shifts and free space propagation phase shifts.
To this end, we propose an iterative algorithm based on block coordinate descent and successive convex approximation techniques. Moreover, we consider the special case with a single pinching antenna, which is a simplified version of the multi-antenna case. Although the formulated problem is still nonconvex, by using the inherent features of the formulated problem, we derive the global optimal solution in closed-form, which offers important insights on the performance of pinching-antenna systems.
Simulation results demonstrate that the pinching-antenna system significantly outperforms conventional fixed-position antenna systems, and the proposed algorithm achieves performance comparable to the computationally intensive exhaustive search based approach.
\end{abstract}

\begin{IEEEkeywords}
     Pinching antenna, flexible-antenna system, non-orthogonal multiple access.
\end{IEEEkeywords}

\section{Introduction} 
Multi-antenna techniques have been viewed as cornerstones to improve the spatial and spectral efficiencies in  wireless communication systems \cite{larsson2014massive,xu2024distributed}. The ability to exploit spatial degrees of freedom for multiplexing and beamforming has enabled unprecedented improvement in throughput and reception reliability. Despite these advantages, conventional multi-antenna techniques predominantly rely on fixed-position antenna arrays, where the antenna elements are statically deployed at predetermined positions. Under this static deployment, the spatial diversity and beamforming gains are constrained by the predetermined array geometry, which may not be optimal in environments with dynamic user distributions and time-varying channel conditions. This motivates the exploration of more flexible antenna configurations. 

Flexible-antenna systems, including reconfigurable intelligent surfaces (RIS) \cite{tang2020wireless,wu2019towards}, fluid antenna systems \cite{wong2020fluid, new2024tutorial}, and movable antenna systems \cite{zhu2023movable, ma2023mimo}, have gained significant attention for their potential to dynamically reconfigure wireless channels and enhance communication performance. By adjusting the reflection coefficients, changing the antenna positions, or modifying the array geometry, these techniques enable adaptive reconfiguration of the wireless environment, thereby improving both the link quality and  spectral efficiency. Despite their advantages, these systems face notable limitations. For instance, RIS-based systems suffer from double attenuation due to the transmitter-RIS link and  RIS-receiver link, which can result in significant path loss. Moreover, fluid and movable antennas are typically constrained to a limited spatial region, e.g., only a few wavelength in extent, thus restricting their ability to combat large-scale path loss or effectively adapt to highly dynamic user distributions.

The pinching-antenna systems can be viewed as a versatile and efficient evolution of flexible-antenna systems to overcome the limitations of conventional systems \cite{ding2024flexible}. A pinching antenna leverages the unique properties of dielectric waveguides to enable flexible and reconfigurable antenna deployment. 
By activating discrete dielectric materials at desired positions along the waveguide, a portion of the propagating radio waves can be induced into these materials, and thus enabling radiation at the selected \textit{pinching} points \cite{suzuki2022pinching}. This mechanism allows antennas to dynamically transmit or receive signals at any position along the waveguide, creating new coverage areas as needed. Unlike conventional antennas, the radiation can be terminated by simply releasing the dielectric material. The ease of installation—simply by attaching separate dielectrics at specific positions—makes pinching antennas cost-effective and adaptable communication solutions for environments demanding flexibility, such as industrial Internet of things (IoT) and dense urban areas. These inherent advantages position pinching antennas as a promising technology for future wireless communication systems.

Inspired by the successful testbed validation of pinching-antenna systems by an industrial pioneer \cite{suzuki2022pinching}, significant research efforts have emerged from the academic community to explore and advance this innovative technology \cite{ding2024flexible,xu2024rate,wang2024antenna,ouyang2025array,tegos2024minimum,yang2025pinching,ding2025blockage,mao2025multi,zhou2025gradient,xu2025pinching}. In particular, a systematic analysis of the ergodic capacity of pinching-antenna systems is provided in \cite{ding2024flexible}, which demonstrates the superiority of pinching-antenna systems over conventional-antenna systems due to their ability to create strong line-of-sight (LoS) links and mitigate large-scale path loss. In \cite{xu2024rate}, a downlink pinching-antenna system is investigated, where multiple pinching antennas are deployed on a waveguide to serve a single-antenna user. A low-complexity, two-stage algorithm is proposed to optimize pinching-antenna positions for maximizing the downlink transmission rate. Meanwhile, the pinching-antenna activation problem is studied in \cite{wang2024antenna}. The formulated problem aims to maximize the sum-rate of downlink users. The array gain of pinching-antenna systems is characterized in \cite{ouyang2025array}. The uplink pinching-antenna system is studied in \cite{tegos2024minimum}, where the minimum of the multiple users' data rates is maximized. 
The impact of LoS blockage on pinching-antenna systems is investigated in \cite{ding2025blockage}. Results show that the LoS blockage can be useful in the multi-user scenario for co-channel interference suppression. 

\begin{figure}[!t]
	\centering
	\includegraphics[width=0.96\linewidth]{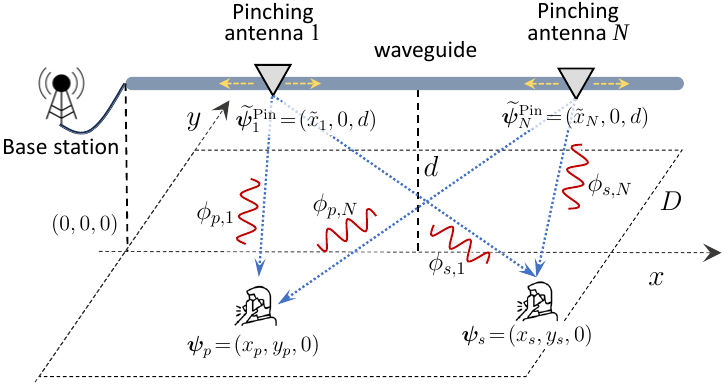}\\
        \captionsetup{justification=justified, singlelinecheck=false, font=small}	
        \caption{An illustration of the considered pinching-antenna system, where $N$ pinching antennas deployed on a waveguide are jointly used to serve two single-antenna users. } \label{fig: system model} 
\end{figure}


In this paper, we investigate a non-orthogonal multiple access (NOMA)-assisted downlink multiuser pinching-antenna system, which includes the problem studied in \cite{xu2024rate} as a special case. In particular, in the considered system, multiple pinching antennas deployed along a waveguide are jointly utilized to serve multiple downlink users, as illustrated in Fig. \ref{fig: system model}. 
The users can either be individually served using conventional orthogonal multiple access (OMA) or jointly served  using NOMA \cite{ding2024flexible,wang2024antenna}. When OMA scheme is used, each user is allocated orthogonal resources, and the problem is reduced to the one studied in \cite{xu2024rate}.   On the other hand, when NOMA scheme is applied, the system spectrum efficiency can be significantly improved by serving multiple users simultaneously; however, it also leads to a much more complex optimization problem, which is the challenge to be addressed in this work. In the literature, there are recent works which investigate the NOMA-assisted pinching antenna design \cite{hu2025sum,xie2025low,zhou2025sum,fu2025power}. Specifically, a multi-waveguide system is considered in \cite{hu2025sum}, where each waveguide is equipped with a pinching antenna. A sum rate maximization problem is formulated by jointly optimizing the beamforming and pinching-antenna position. In \cite{xie2025low} a pinching antenna is utilized to serve multiple users using NOMA. A low-complexity suboptimal pinching-antenna placement algorithm is proposed. In \cite{zhou2025sum}, a NOMA-assisted two-user pinching antenna system is considered. The user power allocation coefficients and the pinching-antenna positions are iteratively updated by using an alternating optimization method. In \cite{fu2025power}, an optimization problem is formulated to minimize the transmit power   in a NOMA-assisted pinching antenna system. Results show that a higher power saving can be achieved when compared with conventional antenna systems.

To obtain better insights on the performance of NOMA-assisted downlink pinching-antenna system,  we focus on a typical two-user NOMA scenario in this paper.
Under the NOMA scheme, the signals of both users are combined using superposition coding. The combined signal is then fed into the waveguide for transmission to the users. We consider a cognitive radio-inspired NOMA (CR-NOMA) system \cite{ding2021no,liu2016nonorthogonal,lv2018cognitive}. Specifically, in a two-user CR-NOMA system, there is a primary user whose quality of service (QoS) requirement needs to be satisfied. The other user is an opportunistic secondary user.  The secondary user aims to maximize its data rate by utilizing the remaining system resources, after the primary user's QoS requirement is satisfied.
In such a CR-NOMA framework, the primary user decodes its information directly. The secondary user first decodes and subtracts the information of the primary user using successive interference cancellation (SIC), and then decodes its own information. We formulate the problem of maximizing the data rate of the secondary user subject to the QoS requirement of the primary user, by jointly optimizing the positions of pinching antennas and the power allocation coefficients.
The formulated problem is nonconvex. Moreover, the positions of the pinching antennas have an impact on the large-scale path loss and two types of phase shifts, namely in-waveguide phase shifts and free space propagation phase shifts, adding complexity to the problem. To this end, we propose efficient algorithms to solve the formulated problem.
The main contributions of this work are summarized as follows:
\begin{itemize}
    \item We begin by addressing the general case where an arbitrary number of pinching antennas are used to serve two users. We formulate a problem to optimize the positions of the pinching antennas and power allocation coefficients. The formulated problem is decoupled into two subproblems: the power allocation subproblem and the pinching-antenna position optimization subproblem. These subproblems are solved iteratively using the block coordinate descent (BCD) method. We note that the power allocation subproblem can be optimally solved when the pinching-antenna positions are fixed. This property implies that the original problem can be globally solved once the optimal pinching-antenna positions are determined, such as using an exhaustive search approach. However, the exhaustive search approach suffers from high computational complexity, especially when the number of pinching antennas is large. To solve the nonconvex pinching-antenna position optimization subproblem efficiently, a successive convex approximation (SCA)-based algorithm is proposed. This leads to a double-loop BCD-SCA algorithm, where the outer loop iteratively updates the power allocation coefficients and pinching-antenna positions, while the inner loop utilizes the SCA-based method to update the pinching-antenna positions. 

    \item We further investigate a special case where only one pinching antenna is available to serve the users. The motivation to study this special case is that it is simpler than the multi-antenna case, and hence it is possible to obtain a closed-form solution which can provide additional  insights. However, we note that the optimization problem for the simplified special case is still nonconvex. We utilize an important feature of the formulated problem and derive a closed-form expression for the global optimal solution. Furthermore, the closed-form solution provides several insightful observations. For example, the larger distance between the user and the waveguide, the closer the pinching antenna will be placed to the user to compensate the large-scale path loss. When both users have the same distance to the waveguide, the pinching antenna will be placed right at the midpoint between the two users on the waveguide.

    \item Extensive numerical results are provided to validate the effectiveness of the proposed NOMA transmission scheme in the pinching-antenna system, along with the algorithm developed to solve the formulated problem. The results demonstrate that the proposed BCD-SCA algorithm converges quickly and achieves superior performance when compared with the high-complexity exhaustive search-based algorithm. Additionally, the pinching-antenna system significantly outperforms the conventional fixed-position antenna system, highlighting its potential as a promising flexible-antenna solution for next-generation wireless communications.
    
\end{itemize}

This paper is organized as follows: Section \ref{sec: system model} introduces the system model and signal model in pinching-antenna system, and the considered optimization problem is formulated. 
In Section \ref{sec: multiple pinching antenna}, we propose a BCD-SCA algorithm to solve the formulated problem for the general multi-antenna case. In Section \ref{sec: single pinching antenna}, we present the special case of the formulated problem where only one pinching antenna is used to serve the users.
Section \ref{sec: simulation} evaluates the performance of the proposed algorithm by numerical simulations. Finally, conclusion is drawn in Section \ref{sec: conclusion}.

 {\bf Notations:} Column vectors are denoted by bold lowercase letters, e.g., $\mathbf{x}$.
$\mathbb{C}^{N}$ denotes the set of $N$-dimensional complex-valued vectors.
The superscript $(\cdot)^\top$ represents the transpose operation.
$\|\mathbf{x}\|^2$ denotes the Euclidean norm of vector $\mathbf{x}$.
$x \sim \mathcal{CN}(a, b)$ denotes that random variable $x$ follows a complex Gaussian distribution with mean $a$ and variance $b$.
$\mathcal{N} \setminus \{1\}$ denotes the set $\mathcal{N}$ excluding the element $1$.
$\max\{a, b\}$ and $\min\{a, b\}$ return the maximum and minimum of $a$ and $b$, respectively.

\section{System Model and Problem Formulation} \label{sec: system model}
Consider a typical two-user NOMA downlink communication scenario, where a base station (BS) is equipped with $N$ pinching antennas on a single waveguide. Let $\mathcal{N} = \{1, \ldots, N\}$ denote the set of pinching antennas. Let $p$ and $s$ denote the primary user and secondary user, respectively. The two users are randomly deployed within a square area of side length $D$, as illustrated in Fig. \ref{fig: system model}. 
The waveguide is positioned parallel to the $x$-axis at a height $d$ and the coordinate of the feed point is denoted by $\widetilde\psib_0^{\pin} = [x_0, 0,d]$.
The free space propagation phase shift of the received signal from the $n$-th pinching antenna to user $m$ is represented by $\phi_{m,n}$, for $m \in \{p,s\}, n \in \mathcal{N}$. The positions of user $m$ and the $n$-th pinching antennas are denoted by $\psib_m = [x_m,y_m,0], m \in \{p,s\},$ and $\widetilde\psib_n^{\pin} = [\tilde x_n, 0,d], n \in \mathcal{N}$, respectively.
The channel vector between the pinching antennas and user $m$ is given by\footnote{In this work, we do not consider the in-waveguide attenuation in the channel model in order to enable simpler derivations and more insightful analysis. This modeling simplification is supported by recent findings in \cite{xu2025pinching}, which show that the resulting performance loss is negligible when the communication area is reasonably bounded. To further support this choice, we also provide numerical comparisons in Section \ref{sec: simulation} to evaluate the impact of in-waveguide attenuation on system performance.}
\begin{align} \label{eqn: h pinching}
    \hb_m = \left[ \frac{\eta^{\frac{1}{2}} e^{-j \frac{2\pi}{\lambda}\|\bm{\psi}_m - \boldsymbol{\widetilde \psi}_1^{\pin}\|}}{\|\boldsymbol{\psi}_m - \boldsymbol{\widetilde \psi}_1^{\pin}\|},..., \frac{\eta^{\frac{1}{2}} e^{-j \frac{2\pi}{\lambda}\|\boldsymbol{\psi}_m - \boldsymbol{\widetilde \psi}_N^{\pin}\|}}{\|\boldsymbol{\psi}_m - \boldsymbol{\widetilde \psi}_N^{\pin}\|}\right]^\top,
\end{align}
where $\eta = \frac{c^2}{16\pi^2 f_c^2}$ is a constant, $c$ denotes the speed of light, $f_c$ is the carrier frequency, and $\lambda$ is the wavelength in free space. Unlike the conventional fixed-position antenna systems, the positions of the pinching antennas are adjustable. Thus, the channel vector $\hb_m$  will change depending on the values of $\tilde x_n$.

\subsection{OMA Transmission Scheme} \label{sec: OMA}
We first consider the OMA transmission scheme. Without loss of generality, time division multiple access (TDMA) scheme is considered, under which the two users are served sequentially in each of the two time slots. The received signal at user  $m \in \{p, s \}$ is given by
\begin{align}
    y_m^{\rm OMA} = \sqrt{\frac{P}{N}}(\hb_m^\pin)^\top \sb_m + w_m, 
\end{align}
where $P$ denotes the total transmit power and $w_m \sim \CN(0,\sigma_m^2)$ is the additive white Gaussian noise at user $m$  with zero mean and variance $\sigma_m^2$.
Here, the transmit power for user $m$  is $\frac{P}{N}$ because it is assumed that the total transmit power $P$ is uniformly allocated to the $N$ pinching antennas. 
On the other hand, since all $N$ pinching antennas are deployed along the same waveguide, the signal transmitted by one antenna is essentially a phase-shifted version of the signal transmitted by another \cite{pozar2021microwave}. Consequently, the signal vector $\sb$ can be represented as follows:
\begin{align} \label{eqn: signal model}
    \sb_m = [e^{-j\theta_1},...,e^{-j\theta_N}]^\top s_m,
\end{align}
where $s_m \in \Cs$ is the signal passed on the waveguide, and $\theta_n = 2\pi\frac{\|\psib_0^{\pin} - \widetilde \psib_n^{\pin}\|}{\lambda_g}$ is the phase shift caused by signal propagation inside the waveguide. Here, $\psib_0^{\pin}$ denotes the position of the feed point of the waveguide, $\lambda_g = \frac{\lambda}{n_{\rm neff}}$ denotes the guided wavelength, and $n_{\rm neff}$ denotes the effective refractive index of a dielectric waveguide \cite{pozar2021microwave}.

Based on the above model, the received signal at user $m \in \{p, s\}$ can be written as 
\begin{align} \label{eqn: received signal}
    y_m^{\rm OMA} =\bigg(\sum\limits_{n=1}^N \frac{\eta^{\frac{1}{2}} e^{-j \left(\frac{2 \pi}{\lambda}\|\boldsymbol{\psi}_m - \tilde{\boldsymbol{\psi}}_n^{\pin}\| +  \theta_n \right)}}{\|\boldsymbol{\psi}_m  -\tilde{\boldsymbol{\psi}}_n^{\pin}\|} \bigg) \sqrt{\frac{P}{N}} s_m + w_m.
\end{align}

Using this model, the achievable data rate of user $m$ for the pinching-antenna system using OMA can be expressed as
\begin{align}
    R_m^{\rm OMA} \!=\! \frac{1}{2}\log \left( 1 \!+\! \bigg| \sum\limits_{n=1}^{N} \frac{\eta^{\frac{1}{2}} e^{-j \left( \frac{2\pi}{\lambda} \| \bm{\psi}_m - \bm{\widetilde \psi}_n^{\pin} \| + \theta_n \right)}}{\| \bm{\psi}_m - \bm{\widetilde\psi}_n^{\pin} \|} \bigg|^2 \!\frac{P}{N \sigma_m^2} \!\right).
\end{align}
Without loss of generality, we assume that the pinching antennas are deployed in a successive order. Then, the sum-rate maximization problem can be formulated as
\begin{subequations} \label{p: original problem}
    \begin{align}
        \underset{\tilde x_1,...,\tilde x_N}{\rm{maximize}} ~&R_p^{\rm OMA} + R_s^{\rm OMA} \\
        \st ~&\tilde x_n - \tilde x_{n-1} \geq \Delta, \forall n\in \Nset \setminus \{ 1 \}, \label{eqn: noncvx antenna spacing}
    \end{align}
\end{subequations}
where constraint \eqref{eqn: noncvx antenna spacing} restricts the antenna spacings to be no smaller than the minimum guide distance, $\Delta$, in order to avoid the antenna coupling.
The rate maximization problem for each user can be decoupled from problem \eqref{p: original problem}, and the algorithm proposed in \cite{xu2024rate} becomes applicable to handle each subproblem. 

\subsection{NOMA Transmission Scheme}
To further improve the spectral efficiency of the pinching-antenna system, one can serve two users simultaneously, which leads to the NOMA transmission scheme. Without loss of generality, we consider a CR-NOMA system \cite{ding2021no,liu2016nonorthogonal,lv2018cognitive}, where the primary user has a QoS requirement that needs to be strictly satisfied. The secondary user is an opportunistic user, who aims to maximize its own data rate by opportunistically utilizing the remaining resources after accommodating the primary user's QoS requirement.
For CR-NOMA, the signals for the two users are first superimposed at the transmitter and then transmitted to them. The superimposed signal is given by
\begin{align}
    s = \sqrt{\alpha_p} s_p + \sqrt{\alpha_s}s_s,
\end{align}
where $\alpha_m \geq 0, m \in \{p,s\},$ are the transmit power coefficients for user $m$, and satisfy $\alpha_p + \alpha_s = 1$. Based on the channel model \eqref{eqn: h pinching} and the signal model \eqref{eqn: signal model}, the effective channel gain at user $m$ can be written as 
\begin{align} \label{eqn: effective channel gain}
    \tilde h_m = \sum_{n=1}^N \frac{\eta^{\frac{1}{2}} e^{-j \big(\frac{2\pi}{\lambda}\|\bm{\psi}_m - \boldsymbol{\widetilde \psi}_n^{\pin}\| + \frac{2\pi}{\lambda_g}\|\psib_0^{\pin} - \widetilde \psib_n^{\pin}\|\big)}}
    {\|\boldsymbol{\psi}_m - \boldsymbol{\widetilde \psi}_n^{\pin}\|}.
\end{align}
Then, the received signal at user $m \in \{p,s\}$ is given by
\begin{align}
    &y_m^{\rm NOMA} = w_m +\notag\\
    &\bigg(\sum\limits_{n=1}^N \frac{\eta^{\frac{1}{2}} e^{-j \big(\frac{2\pi}{\lambda}\|\bm{\psi}_m - \boldsymbol{\widetilde \psi}_n^{\pin}\| + \frac{2\pi}{\lambda_g}\|\psib_0^{\pin} - \widetilde \psib_n^{\pin}\|\big)}}{\|\boldsymbol{\psi}_m  -\tilde{\boldsymbol{\psi}}_n^{\pin}\|} \bigg) \sqrt{\frac{P}{N}} s.
\end{align}

In CR-NOMA system, the secondary user first performs SIC to decode and remove the primary user's information before decoding its own. The associated signal-to-interference-plus-noise ratio (SINR) and signal-to-noise ratio (SNR) for the two SIC steps are respectively given by
\begin{subequations}
    \begin{align}
        \SINR_s^{(s_p)} &= \frac{\alpha_p P |\tilde h_s|^2}{\alpha_s P |\tilde h_s|^2 + N\sigma_s^2},\\
        \SNR_s^{(s_s)} &= \frac{\alpha_s P |\tilde h_s|^2}{N\sigma_s^2}.
    \end{align}
\end{subequations}
The primary user directly decodes its own information, $s_p$, and the associated SINR can be written as 
\begin{align}
    \SINR_p^{(s_p)} = \frac{\alpha_p P |\tilde h_p|^2}{\alpha_s P |\tilde h_p|^2 + N\sigma_p^2}.
\end{align}

In our problem formulation, the objective is to maximize the data rate of the secondary user, while guaranteeing the data rate requirement of the primary user. The optimization problem can be formulated as 
\begin{subequations} \label{p: rate maximization origin}
    \begin{align}
        \underset{\substack{\alpha_p,\alpha_s, {\tilde x_1,...,\tilde x_N}}}{\rm{maximize}} ~ &\frac{\alpha_s P |\tilde h_s|^2}{N\sigma_s^2} \\
        \st ~& \frac{\alpha_p P |\tilde h_p|^2}{\alpha_s P |\tilde h_p|^2 + N\sigma_p^2} \geq \gamma_p, \label{eqn: rate maximization origin 1}\\
        & \frac{\alpha_p P |\tilde h_s|^2}{\alpha_s P |\tilde h_s|^2 + N\sigma_s^2} \geq \gamma_p, \label{eqn: rate maximization origin 2}\\
        & \tilde x_n - \tilde x_{n-1} \geq \Delta, \hspace{3mm}\forall n\in \Nset \setminus \{ 1 \}, \label{eqn: rate maximization origin 3}\\
        & \alpha_p + \alpha_s = 1, \alpha_p\geq 0, \alpha_s\geq 0, \label{eqn: rate maximization origin 4}
    \end{align}
\end{subequations}
where $\gamma_p$ denotes the primary user's target SINR.
Constraint \eqref{eqn: rate maximization origin 1} corresponds to the data rate requirement of the primary user.
Constraint \eqref{eqn: rate maximization origin 2} guarantees the primary user's information can be successfully removed at the secondary user by the SIC operation. The minimum antenna spacing constraint is given by constraint \eqref{eqn: rate maximization origin 3}. Constraint \eqref{eqn: rate maximization origin 4} corresponds to the constraint on the power allocation coefficients. 
Problem \eqref{p: rate maximization origin} is nonconvex and challenging to solve due to those coupling optimization variables.

\section{Proposed Algorithm for Solving Problem \eqref{p: rate maximization origin}}  \label{sec: multiple pinching antenna}
In this section, we present an algorithm to solve problem \eqref{p: rate maximization origin} for the general case with an arbitrary number of pinching antennas. Problem \eqref{p: rate maximization origin} is difficult to solve due to the coupled nature of the power allocation coefficients with the users' effective channel gains, as well as the complex dependence of these channel gains on the pinching-antenna position variables. 
To solve problem \eqref{p: rate maximization origin} in an efficient manner, we apply a BCD-based algorithm, which allows us to decompose the problem into subproblems and solve them iteratively. Specifically, we decompose problem \eqref{p: rate maximization origin} into two subproblems: the first subproblem aims to optimize the power allocation coefficients while keeping the pinching-antenna positions to be fixed; the second subproblem aims to optimize the pinching-antenna positions with the power allocation coefficients held constant.

\subsection{Optimize the Power Allocation Coefficients}\label{sec: update of power allocation}
In this subsection, we aim to determine the power allocation coefficients by considering the pinching-antenna positions to be fixed.  From constraint \eqref{eqn: rate maximization origin 4}, we have $\alpha_s = 1- \alpha_p$. The associated optimization problem can be formulated as
\begin{subequations}
    \begin{align}
        \underset{\substack{0 \leq \alpha_s \leq 1}}{\rm{maximize}} ~ &\frac{\alpha_s P |\tilde h_s|^2}{N\sigma_s^2}  \label{eqn: power allocation obj}\\
        \st~&  (P |\tilde h_p|^2 + P \gamma_p |\tilde h_p|^2) \alpha_s  \leq P |\tilde h_p|^2 - N \sigma_p^2 \gamma_p, \label{eqn: power allocation c1}\\
        &  (P |\tilde h_s|^2 + P \gamma_p |\tilde h_s|^2) \alpha_s  \leq P |\tilde h_s|^2 - N \sigma_s^2 \gamma_p. \label{eqn: power allocation c2}
    \end{align}
\end{subequations}
From constraints \eqref{eqn: power allocation c1} and \eqref{eqn: power allocation c2}, we have 
\begin{align}
    \alpha_s \leq A,
\end{align}
where 
\begin{align}
    A \triangleq \min \left\{\frac{ P |\tilde h_p|^2 - N \sigma_p^2 \gamma_p}{P |\tilde h_p|^2 + P \gamma_p |\tilde h_p|^2},\frac{ P |\tilde h_s|^2 - N \sigma_s^2 \gamma_p}{P |\tilde h_s|^2 + P \gamma_p |\tilde h_s|^2}\right\}.
\end{align}
Since the objective function \eqref{eqn: power allocation obj} is monotonically increasing with $\alpha_s$, the optimal power allocation coefficients are given by
\begin{subequations} \label{eqn: power allocation}
    \begin{align}
        \alpha_s^* &\!=\! \max\{0,A\}, \\
        \alpha_p^* &= 1 - \alpha_s^*.
    \end{align}
\end{subequations}
Note that if the optimal pinching-antenna positions are available, for example, through an exhaustive search based method, the global optimal solution to problem \eqref{p: rate maximization origin} can be found as in \eqref{eqn: power allocation}. 
However, the exhaustive search approach is computationally expensive, particularly when the number of pinching antennas is large. 
This issue motivates the development of a low-complexity suboptimal algorithm aiming to obtain a set of desirable pinching-antenna positions in an efficient way. 
The performance of the proposed low-complexity algorithm will be evaluated against the optimal solution obtained through a high-complexity exhaustive search in Section \ref{sec: simulation}.

\subsection{Optimize the Pinching-Antenna Positions}
In this subsection, we aim to optimize the pinching-antenna positions with the obtained power allocation coefficients in the previous subsection. 
First, by letting $C_m = y_m^2 + d^2, {\textrm{for}}~ m \in \{p,s\}$, and using the fact that 
    \begin{align}
     &\|\bm{\psi}_m - \boldsymbol{\widetilde \psi}_n^{\pin}\| = \sqrt{(\tilde x_n - x_m)^2 + C_m}, \  \   m \in \{p,s\}, n \in \mathcal{N}, \\
     &\|\psib_0^{\pin} - \widetilde \psib_n^{\pin}\| = \tilde x_n - x_0, \  \   n \in \mathcal{N},
    \end{align}
the channel coefficients can be represented by
\begin{align} \label{eqn: effective channel gain 2}
    \tilde h_m = \sum_{n=1}^N \frac{\eta^{\frac{1}{2}} e^{-j \big(\frac{2\pi}{\lambda}\sqrt{(\tilde x_n - x_m)^2 + C_m} + \frac{2\pi}{\lambda_g}(\tilde x_n - x_0)\big)}}
    {\sqrt{(\tilde x_n - x_m)^2 + C_m}}, \notag \\
    \  \ \ \    m \in \{p, s\}.
\end{align}
Then, we introduce variables $d_{m,n}, t_{m,n}, {\textrm{for}}~ m \in \{p, s\}, n \in \Nset$, and let
\begin{align}
    d_{m,n} &= \sqrt{(\tilde x_n - x_m)^2 + C_m}, \ \ \ m \in \{p, s\}, n \in \Nset,\\
    t_{m,n} &= \frac{\eta^{\frac{1}{2}} e^{-j \big(\frac{2\pi}{\lambda}d_{m,n} + \frac{2\pi}{\lambda_g}(\tilde x_n - x_0)\big)}} {d_{m,n}}, \ \ \ m \in \{p, s\}, n \in \Nset.
\end{align}
Furthermore, $t_{m,n} \in \mathbb{C}$ can be represented as $t_{m,n} = t_{m,n}^{\mathcal{R}} + j t_{m,n}^{\mathcal{I}}$, where $t_{m,n}^{\mathcal{R}}$ and $t_{m,n}^{\mathcal{I}}$ denote the real part and imaginary part of $t_{m,n}$, respectively. In particular, $t_{m,n}^{\mathcal{R}}$ and $t_{m,n}^{\mathcal{I}}$ are respectively given by 
\begin{subequations}
    \begin{align}
        t_{m,n}^{\mathcal{R}} &= \frac{\sqrt{\eta}}{d_{m,n}} \cos\Big(\frac{2\pi}{\lambda}d_{m,n} + \frac{2\pi}{\lambda_g}(\tilde x_n - x_0) \Big),  \\
        t_{m,n}^{\mathcal{I}} &= \frac{\sqrt{\eta}}{d_{m,n}} \sin \Big(\frac{2\pi}{\lambda}d_{m,n} + \frac{2\pi}{\lambda_g}(\tilde x_n - x_0) \Big).
    \end{align}
\end{subequations}
Then, the complex-valued effective channels can be written as
\begin{align}
    \tilde h_m &= \tilde h_m^{\mathcal{R}} + j \tilde h_m^{\mathcal{I}}\notag\\
     &=\sum_{n=1}^N t_{m,n}^{\mathcal{R}} + j \sum_{n=1}^Nt_{m,n}^{\mathcal{I}},
\end{align}
where $\tilde h_m^{\mathcal{R}}$ and $\tilde h_m^{\mathcal{I}}$ denote the real part and imaginary part of $\tilde h_m, \textrm{for}~ m \in \{p,s\}$, respectively.

With the above definitions, the pinching-antenna position optimization problem can be written as
\begin{subequations} \label{p: rate maximization transformation 1}
    \begin{align}
        \underset{\substack{d_{m,n},\tilde h_m^{\mathcal{R}},\tilde h_m^{\mathcal{I}},\\ \tilde x_n,t_{m,n}^{\mathcal{R}},t_{m,n}^{\mathcal{I}}}}{\rm{maximize}} ~ &\frac{\alpha_s P \big((h_s^{\mathcal{R}})^2 + (h_s^{\mathcal{I}})^2\big)}{N\sigma_s^2} \\
        \st ~& (\alpha_pP - \alpha_s P \gamma_p) \big((h_p^{\mathcal{R}})^2 + (h_p^{\mathcal{I}})^2\big) \geq N \sigma_p^2 \gamma_p, \label{eqn: rate maximization t11}\\
        & (\alpha_pP - \alpha_s P \gamma_p) \big((h_s^{\mathcal{R}})^2 + (h_s^{\mathcal{I}})^2\big) \geq N \sigma_s^2 \gamma_p, \label{eqn: rate maximization t12}\\
        &d_{m,n} = \sqrt{(\tilde x_n - x_m)^2 + C_m}, \notag \\
        &\quad\quad\quad\quad\quad \quad \quad  \  \  \  m \in \{p, s\}, n \in \Nset, \label{eqn: rate maximization t13}\\
        &t_{m,n}^{\mathcal{R}} = \frac{\sqrt{\eta}}{d_{m,n}} \cos\Big(\frac{2\pi}{\lambda}d_{m,n} + \frac{2\pi}{\lambda_g}(\tilde x_n - x_0) \Big),  \notag\\
        &\quad\quad\quad\quad\quad \quad \quad  \  \  \  m \in \{p,s\}, n \in \mathcal{N}, \label{eqn: rate maximization t14}\\
        &t_{m,n}^{\mathcal{I}} = \frac{\sqrt{\eta}}{d_{m,n}} \sin \Big(\frac{2\pi}{\lambda}d_{m,n} + \frac{2\pi}{\lambda_g}(\tilde x_n - x_0) \Big),  \notag\\
        &\quad\quad\quad\quad\quad \quad \quad  \  \  \  m \in \{p,s\}, n \in \mathcal{N}, \label{eqn: rate maximization t15}\\
        & \tilde x_n - \tilde x_{n-1} \geq \Delta, \ \ \ n \in \mathcal{N}  \setminus \{ 1 \}, \label{eqn: rate maximization t16}  \\
        & \tilde h_m^{\mathcal{R}} = \sum_{n=1}^N t_{m,n}^{\mathcal{R}}, \ \  \tilde h_m^{\mathcal{I}} = \sum_{n=1}^N t_{m,n}^{\mathcal{I}}, \label{eqn: rate maximization t17}\notag\\
        &\quad\quad\quad\quad\quad \quad \quad  \  \  \  m \in \{p, s\}, n \in \mathcal{N}.
    \end{align}
\end{subequations}
It is noted from constraints \eqref{eqn: rate maximization t11} and \eqref{eqn: rate maximization t12} that, if $\alpha_pP - \alpha_s P \gamma_p = 0$,  problem \eqref{p: rate maximization transformation 1} becomes infeasible. 
On the other hand, when $\alpha_pP - \alpha_s P \gamma_p < 0$, 
constraints \eqref{eqn: rate maximization t11} and \eqref{eqn: rate maximization t12} become convex second-order cone constraints. 
Conversely, if $\alpha_pP - \alpha_s P \gamma_p > 0$, constraints \eqref{eqn: rate maximization t11} and \eqref{eqn: rate maximization t12} will be nonconvex. The  nonconvexities of problem \eqref{p: rate maximization transformation 1} are also due to the objective function and constraints \eqref{eqn: rate maximization t13}, \eqref{eqn: rate maximization t14}, and \eqref{eqn: rate maximization t15}, where the optimization variables are coupled and intricately related to the pinching-antenna position variables. 
Therefore, problem \eqref{p: rate maximization transformation 1} is still complicated and nonconvex, making it challenging to solve. However, compared to the original problem \eqref{p: rate maximization origin}, it is structured in a way that makes systematic approaches applicable. In what follows, we will resort to the SCA-based algorithm to solve it in an iterative fashion.

First,  by introducing an auxiliary variable $z$ and using the epigraph reformulation, the objective function can be rewritten as
\begin{subequations}
    \begin{align}
        \underset{\substack{z, \tilde h_{s}^{\mathcal{R}}, \tilde h_{s}^{\mathcal{I}}}}{\rm{maximize}}~& z \\
        \st ~&  \alpha_s P \big((\tilde h_{s}^{\mathcal{R}})^2 + (\tilde h_{s}^{\mathcal{I}})^2\big) \geq N \sigma_s^2 z. \label{p: interested rate maximization 1}
    \end{align}
\end{subequations}
By defining $\gb_m = [\tilde{h}_m^{\mathcal{R}},\tilde{h}_m^{\mathcal{I}}]^\top, m \in \{p,s\}$, constraint \eqref{p: interested rate maximization 1} can be rewritten as
\begin{align} 
     \|\gb_s\|^2 \geq \frac{N\sigma_s^2}{\alpha_sP} z.
\end{align}
Then, applying the first-order Taylor approximation to $\|\gb_s\|^2$, this constraint can be approximately represented by
\begin{align} \label{eqn: sca 1}
    \|\gb_s^k\|^2 + 2\|\gb_s^k\|^\top(\gb_s - \gb_s^k)  \geq \frac{N\sigma_s^2}{\alpha_sP} z,
\end{align}
where $\gb_s^k = [(\tilde{h}_s^{\mathcal{R}})^k,(\tilde{h}_s^{\mathcal{I}})^k]^\top$ denotes the obtained 
solution of $\gb_s$ in the last iteration (i.e., the $k$-th iteration). The above steps can be similarly applied to \eqref{eqn: rate maximization t11} and \eqref{eqn: rate maximization t12} if $\alpha_pP - \alpha_s P \gamma_p > 0$. Specifically, these constraints can be approximately represented by
\begin{subequations}\label{eqn: sca 2}
    \begin{align}
        \|\gb_p^k\|^2 + 2\|\gb_p^k\|^\top(\gb_p - \gb_p^k)  \geq \frac{N\sigma_p^2\gamma_p}{\alpha_pP - \alpha_s P \gamma_p}, \label{eqn: qos approximation 1}\\
        \|\gb_s^k\|^2 + 2\|\gb_s^k\|^\top(\gb_s - \gb_s^k)  \geq \frac{N\sigma_s^2\gamma_p}{\alpha_pP - \alpha_s P \gamma_p}, \label{eqn: qos approximation 2}
    \end{align}
\end{subequations}
where $\gb_p^k = [(\tilde{h}_p^{\mathcal{R}})^k,(\tilde{h}_p^{\mathcal{I}})^k]^\top$ denotes the obtained solution of $\gb_p$ in the $k$-th iteration.

For constraint \eqref{eqn: rate maximization t13}, one can approximate it by
\begin{align} \label{eqn: d approximation}
    d_{m,n} = d_{m,n}^k + (d_{m,n}^k)' (\tilde x_n - \tilde x_n^k) \ \ \   m \in \{p, s\}, n \in \mathcal{N},
\end{align}
where $(d_{m,n}^k)'$ represents the first-order derivative of $d_{m,n}$ evaluated at $\tilde x_n = \tilde x_n^k$ and can be written as
\begin{align}
    (d_{m,n}^k)' = \frac{\tilde x_n^k - x_m}{\big((\tilde x_n^k - x_m)^2 + C_m\big)}.
\end{align}

Constraints \eqref{eqn: rate maximization t14} and \eqref{eqn: rate maximization t15} can be handled in a similar way, which are respectively approximated by
\begin{subequations} \label{eqn: t approximation}
    \begin{align}
        t_{m,n}^{\mathcal{R}} &= t_{m,n}^{\mathcal{R},k} + (t_{m,n}^{\mathcal{R},k})' (\tilde x_n \!-\! \tilde x_n^k), \ \ \   m \in \{p, s\}, n \in \mathcal{N},\\
        t_{m,n}^{\mathcal{I}} &= t_{m,n}^{\mathcal{I},k} + (t_{m,n}^{\mathcal{I},k})' (\tilde x_n \!-\! \tilde x_n^k), \ \ \   m \in \{p, s\}, n \in \mathcal{N},
    \end{align}
\end{subequations}
where $t_{m,n}^{\mathcal{R},k}$ and $t_{m,n}^{\mathcal{I},k}$ denote the values $t_{m,n}^{\mathcal{R}}$ and $t_{m,n}^{\mathcal{I}}$ obtained at the 
$k$-th iteration, respectively. Similarly, $(t_{m,n}^{\mathcal{R},k})'$ and $(t_{m,n}^{\mathcal{I},k})'$ denote the first-order derivatives of $t_{m,n}^{\mathcal{R}}$ and $t_{m,n}^{\mathcal{I}}$ evaluated at $\tilde x_n = \tilde x_n^k$ and are given by
\begin{subequations}\label{p: bcd-sca}
    \begin{align}
        (t_{m,n}^{\mathcal{R},k})' &= -\frac{2\pi\sqrt{\eta}}{\lambda_g d_{m,n}^k} \sin\Big(\frac{2\pi}{\lambda}d_{m,n}^k \!+\! \frac{2\pi}{\lambda_g}(\tilde x_n^k \!-\! x_0) \Big), \notag\\
        &\quad\quad\quad\quad\quad  \qquad\qquad   \  \  \  m \in \{p, s\}, n \in \mathcal{N},\\
        (t_{m,n}^{\mathcal{I},k})' &= \frac{2\pi\sqrt{\eta}}{\lambda_g d_{m,n}^k} \cos\Big(\frac{2\pi}{\lambda}d_{m,n}^k + \frac{2\pi}{\lambda_g}(\tilde x_n^k - x_0) \Big),\notag\\
        &\quad\quad\quad\quad\quad  \qquad\qquad   \  \  \  m \in \{p, s\}, n \in \mathcal{N}.
    \end{align}
\end{subequations}

Based on the above approximations, the approximated convex problem to be solved by the SCA-based algorithm in the $(k+1)$-th iteration is given by
    \begin{align} \label{p: sca}
    \underset{\substack{z, d_{m,n},\tilde h_m^{\mathcal{R}},\tilde h_m^{\mathcal{I}},\\ \tilde x_n,t_{m,n}^{\mathcal{R}},t_{m,n}^{\mathcal{I}}}}{\rm{maximize}}~&   z \\
    \st~& \textrm{Constraints}~\eqref{eqn: rate maximization t11},\eqref{eqn: rate maximization t12}, \eqref{eqn: rate maximization t16},\notag\\
    & \eqref{eqn: rate maximization t17}, \eqref{eqn: sca 1}, \eqref{eqn: d approximation}, \textrm{and}~ \eqref{eqn: t approximation}. \notag
    \end{align}
Here, it is assumed that $\alpha_pP - \alpha_s P \gamma_p < 0$. Otherwise, one needs to replace constraints \eqref{eqn: rate maximization t11} and \eqref{eqn: rate maximization t12} by constraint \eqref{eqn: sca 2}. 
The SCA-based algorithm iteratively solves problem \eqref{p: sca} until a certain convergence criterion is satisfied.
We note that the SCA-based algorithm guarantees to converge to a stationary point of problem \eqref{p: rate maximization transformation 1} \cite{li2012coordinated}. After the algorithm has converged, the obtained channel coefficients can be substituted into \eqref{eqn: power allocation} to update the power allocation coefficients for the next iteration.

\subsubsection{Initialization of the SCA-based Algorithm} We now present how to find an efficient initial point for the SCA-based algorithm. It is noticed that, given a set of initial pinching-antenna positions, i.e., $\tilde x_1^0, \ldots, \tilde x_N^0$, the initial values of other variables can be determined accordingly. Therefore, we need to find a set of pinching-antenna positions to initialize the SCA-based algorithm. 

Recall that our objective is to maximize the data rate of the secondary user while satisfying the data rate requirement of the primary user. 
Since the user data rates are positively related with the users' effective channel gains, we are interested in finding a set of pinching-antenna positions to maximize the users' effective channel gains.
However, it is not a trivial problem, because the positions of the pinching antennas affect both the path losses and the phase shifts of the users' effective channel gains. 
On the other hand, an important insight from \cite{ding2024flexible,xu2024rate} is that, in order to maximize the data rate, it is critical to minimize the path losses between the pinching antennas and the user, while the phase shifts can be efficiently modified by adjusting the pinching-antenna positions on the scale of wavelength.  
Based on this insight, we consider a similar problem as that in \cite{xu2024rate}, which aims at minimizing the effects from path losses on the effective channel gains of the two users. The associated problem is given by
\begin{align} \label{p: initialization}
    \underset{\tilde x_1,...,\tilde x_N}{\rm{maximize}} ~&  \sum_{n=1}^N \underbrace{\big[(\tilde x_n \!-\! x_p)^2 \!+\! C_p \big]^{-\frac{1}{2}}}_{f_p(\tilde x_n)} + \underbrace{\big[(\tilde x_n \!-\! x_s)^2 \!+\! C_s \big]^{-\frac{1}{2}}}_{f_s(\tilde x_n)},
\end{align}
where $f_p(\tilde x_n)$ and $f_s(\tilde x_n)$ characterize the effects of path losses on the effective channel gains between the $n$-th pinching antenna and the primary user and the secondary user, respectively. 
To avoid antenna coupling and facilitate the initialization, we set $\tilde x_n - \tilde x_{n-1} = \Delta,\ \textrm{for}~ n \in \Nset \setminus 1$, as an approximation of the original spacing constraint in \eqref{eqn: rate maximization t16}. This equality constraint is used solely during initialization, while the original inequality form in \eqref{eqn: rate maximization t16} is strictly enforced during the BCD-SCA algorithm to solve problem \eqref{p: bcd-sca}.
Then, problem \eqref{p: initialization} can be rewritten as
\begin{align} \label{p: initialization x1}
    \underset{\tilde x_1}{\rm{maximize}} ~ & \sum_{n=1}^N \big[(\tilde x_1 + (n-1)\Delta - x_p)^2 + C_p \big]^{-\frac{1}{2}} \notag\\
    &+ \sum_{n=1}^N \big[(\tilde x_1 + (n-1)\Delta - x_s)^2 + C_s \big]^{-\frac{1}{2}}.
\end{align}
Let $g(\tilde x_1)$ denote the objective function of problem \eqref{p: initialization x1}. As seen, $g(\tilde x_1)$ is a function of $\tilde x_1$, making it challenging to obtain the optimal solution of problem \eqref{p: initialization x1}. 
Observing that $g(\tilde x_1)$ is smooth and twice differentiable, which motivates us to use the Newton's method to obtain a local maximum in an efficient way \cite{nocedal2006numerical}. In particular, the Newton's method attempts to find a critical point of $g(\tilde x_1)$ in an iterative manner, and the process is repeated as
\begin{align} \label{eqn: newton iteration}
    \tilde x_1^{k+1} = \tilde x_1^{k} - \frac{g'(\tilde x_1^{k})}{g''(\tilde x_1^{r})}, k = 1,2,...,
\end{align}
$g'(\tilde x_1^{k})$ and $g''(\tilde x_1^{k})$ respectively denote the first- and second-order derivatives of $f(\tilde x_1)$ at point $\tilde x_1^{k}$ and are calculated by
\begin{align}
    g'(\tilde x_1) &=  -\sum_{n=1}^N \frac{\left( \tilde{x}_1 + (n-1) \Delta - x_p \right)}{\left[ \left( \tilde{x}_1 + (n-1) \Delta - x_p \right)^2 + C_p \right]^{\frac{3}{2}}} \notag\\
    &\quad - \sum_{n=1}^N \frac{\left( \tilde{x}_1 + (n-1) \Delta - x_s \right)}{\left[ \left( \tilde{x}_1 + (n-1) \Delta - x_s \right)^2 + C_s \right]^{\frac{3}{2}}},\\
    g''(\tilde x_1) &= \sum_{n=1}^N \frac{3 \left( \tilde{x}_1 + (n-1) \Delta - x_p \right)^2}{\left[ \left( \tilde{x}_1 + (n-1) \Delta - x_p \right)^2 + C_p \right]^{\frac{5}{2}}}  \notag\\
    &\quad - \sum_{n=1}^N \left[ \left( \tilde{x}_1 + (n-1) \Delta - x_p \right)^2 + C_p \right]^{-\frac{3}{2}}  \notag\\
    &\quad + \sum_{n=1}^N \frac{3 \left( \tilde{x}_1 + (n-1) \Delta - x_s \right)^2}{\left[ \left( \tilde{x}_1 + (n-1) \Delta - x_s \right)^2 + C_s \right]^{\frac{5}{2}}}   \notag\\
    &\quad - \sum_{n=1}^N \left[ \left( \tilde{x}_1 + (n-1) \Delta - x_s \right)^2 + C_s \right]^{-\frac{3}{2}}.
\end{align}
It is noted that the convergence property of the Newton's method depends on the shape of $g(\tilde x_1)$ and it may converge to a local minimum of $g(\tilde x_1)$, where $g''(\tilde x_1) > 0$. In this case, we need to choose another initial point of $\tilde x_1$ and repeat the iterations in \eqref{eqn: newton iteration}.
However, we would like to note that our numerical results show that function $g(\tilde x_1)$ is  unimodal with respect to $\tilde x_1$. This property implies that the obtained solution by the Newton's method is the maximum of $g(\tilde x_1)$. 

\begin{algorithm}[t] \small
	\caption{Proposed BCD-SCA algorithm for solving problem \eqref{p: rate maximization origin}}
	\label{alg: BCD-SCA}
	\begin{algorithmic}[1]
		\STATE Given $C_p$, $C_s$ and $\Delta$, run the Newton's method to obtain a set of initial pinching-antenna positions.
		\WHILE{a certain convergence criteria is not satisfied} 
                \WHILE{a certain convergence criteria is not satisfied} 
                \STATE Solve problem \eqref{p: sca} to obtain a set of pinching-antenna positions.
                \ENDWHILE
		\STATE With the obtained pinching-antenna positions, update the power allocation coefficients based on \eqref{eqn: power allocation}
		\ENDWHILE
		\STATE \textbf{Output}: The pinching-antenna positions $\tilde x_1^*,...,\tilde x_N^*$, and the power allocation coefficients $\alpha_p^*$ and $\alpha_s^*$.
	\end{algorithmic} 
\end{algorithm}

The proposed BCD-SCA algorithm iteratively updates the power allocation coefficients based on \eqref{eqn: power allocation} and updates the pinching-antenna positions by solving problem \eqref{p: sca}, until a certain convergence criterion is satisfied.
The detailed procedure of the proposed BCD-SCA algorithm is summarized in Algorithm \ref{alg: BCD-SCA}.


\section{Optimal Solution for the Single-Pinching-Antenna Case} \label{sec: single pinching antenna}
In this section, we consider a special case, where there is only one pinching antenna on the waveguide for data transmission, i.e., $N=1$. In this case, the channel for user $m$ reduces to 
\begin{align} \label{eqn: channel n1}
    \tilde h_m = \frac{\eta^{\frac{1}{2}} e^{-j \big(\frac{2\pi}{\lambda}\|\bm{\psi}_m - \boldsymbol{\widetilde \psi}^{\pin}\| + \frac{2\pi}{\lambda_g}\|\psib_0^{\pin} - \widetilde \psib^{\pin}\|\big)}}
    {\|\boldsymbol{\psi}_m - \boldsymbol{\widetilde \psi}^{\pin}\|},  m \in \{p,s\},
\end{align}
where $\boldsymbol{\widetilde \psi}^{\pin} = [\tilde x, 0, d]$ denotes the position of the pinching antenna. With \eqref{eqn: channel n1}, we have 
\begin{align}
    |\tilde h_m|^2 = \frac{\eta}{(\tilde x - x_m)^2 + C_m}, m \in \{p,s\}.
\end{align} 
Consequently, the optimization problem can be simplified as 
\begin{subequations} \label{p: rate maximization special 1}
    \begin{align}
        \underset{\alpha_p,\alpha_s, x}{\rm{maximize}} ~& \frac{\alpha_s P\eta}{\big((\tilde x - x_s)^2 + C_s \big)\sigma_s^2} \label{eqn: rate sa obj}\\
        \st ~& \frac{\alpha_p P \eta}{\alpha_s P \eta + \big((\tilde x - x_p)^2 + C_p\big)\sigma_p^2} \geq \gamma_p, \label{eqn: rate maximization special 1}\\
        & \frac{\alpha_p P \eta}{\alpha_s P \eta + \big((\tilde x - x_s)^2 + C_s\big)\sigma_s^2} \geq \gamma_p, \label{eqn: rate maximization special 2}\\
        & \alpha_p + \alpha_s = 1, \alpha_p \geq 0, \alpha_s \geq 0.  \label{eqn: rate maximization special 3}
    \end{align}
\end{subequations}
By substituting $\alpha_s = 1 - \alpha_p$ into problem \eqref{p: rate maximization special 1} and with some algebraic manipulations, problem \eqref{p: rate maximization special 1} can be further rewritten as 
\begin{subequations} \label{p: rate maximization special 2}
    \begin{align}
        \underset{x, 0 \leq \alpha_p \leq 1}{\rm{maximize}} ~& \frac{(1-\alpha_p) P\eta}{\big((x - x_s)^2 + C_s \big)\sigma_s^2}\\
        \st ~& (P\eta + P\eta\gamma_p) \alpha_p - (\tilde x - x_p)^2 \sigma_p^2\gamma_p \geq \notag\\
        &\qquad\qquad\qquad\qquad~ P\eta\gamma_p + C_p\sigma_p^2 \gamma_p, \label{eqn: rate maximization special 21}\\
        & (P\eta + P\eta\gamma_p) \alpha_p - (\tilde x - x_s)^2 \sigma_s^2\gamma_p \geq \notag\\
        &\qquad\qquad\qquad\qquad~  P\eta\gamma_p + C_s\sigma_s^2 \gamma_p. \label{eqn: rate maximization special 22}
    \end{align}
\end{subequations}
Although problem \eqref{p: rate maximization special 2} is significantly simplified compared to problem \eqref{p: rate maximization origin}, it is still nonconvex and challenging to solve. 
Problem \eqref{p: rate maximization special 2} can be solved by using an approximation-based algorithm, which leads to a suboptimal solution. In the following, we propose a low-complexity algorithm which utilizes the feature of the formulated optimization problem and yields an insightful optimal solution.
Before solving problem \eqref{p: rate maximization special 2}, we present the following lemma.

\begin{lem} \label{lem: location of x}
Suppose that problem \eqref{p: rate maximization special 2} is feasible, the optimal $x$ satisfies the following condition:
\begin{align}
     x_{\min} \leq \tilde x^* \leq x_{\max},
    \end{align}
    where $ x_{\min} = \min\{x_p,x_s\}$ and $x_{\max} = \max\{x_p, x_s\}$.
\end{lem}

\textit{Proof:} See Appendix \ref{appd: lemma location of x}. \hfill $\blacksquare$

Next, we present the feasibility condition of problem \eqref{p: rate maximization special 2}, which is described in the following lemma.
\begin{lem} \label{lem: feasibility}
    Problem \eqref{p: rate maximization special 2} is feasible if and only if 
    \begin{align} \label{eqn: feasiblity condition}
        P\eta \geq \max\{C_p\sigma_p^2 \gamma_p,C_s\sigma_s^2 \gamma_p\}.
    \end{align}
\end{lem}

\textit{Proof:}
First, based on \eqref{eqn: rate maximization special 21} and \eqref{eqn: rate maximization special 22}, we have
\begin{align} 
    (\tilde x - x_p)^2 &\leq \frac{(P\eta + P\eta\gamma_p) \alpha_p - P\eta\gamma_p - C_p\sigma_p^2 \gamma_p}{\sigma_p^2\gamma_p}.\label{eqn: relationship of optimal solutions11} \\
    (\tilde x - x_s)^2 &\leq \frac{(P\eta + P\eta\gamma_p) \alpha_p - P\eta\gamma_p - C_s\sigma_s^2 \gamma_p}{\sigma_s^2\gamma_p}.\label{eqn: relationship of optimal solutions12}
\end{align}
From \eqref{eqn: relationship of optimal solutions11} and \eqref{eqn: relationship of optimal solutions12}, one can see that problem \eqref{p: rate maximization special 2} is feasible if and only if 
\begin{align}
    &(P\eta + P\eta\gamma_p) \alpha_p \notag\\
    &\geq P\eta\gamma_p + \max\{C_p\sigma_p^2 \gamma_p,C_s\sigma_s^2 \gamma_p\}, \exists \alpha_p \in [0,1].
\end{align}
Since $P\eta + P\eta\gamma_p > 0$, the feasibility condition can be simplified as $P\eta \geq \max\{C_p\sigma_p^2 \gamma_p,C_s\sigma_s^2 \gamma_p\}$, i.e., \eqref{eqn: feasiblity condition}. The proof is completed. \hfill $\blacksquare$

From the proof of Lemma \ref{lem: feasibility}, we also have the following conclusions. When $P\eta = C_p\sigma_p^2 \gamma_p \geq C_s\sigma_s^2 \gamma_p$, the optimal solution is $\tilde x^* = x_p$. On the other hand, when $P\eta = C_s\sigma_s^2 \gamma_p \geq C_p\sigma_p^2 \gamma_p$, the optimal solution is $\tilde x^* = x_s$. In these two cases, all the transmission power is allocated to the primary user, i.e., $\alpha_p = 1$, with which one of constraints \eqref{eqn: rate maximization special 21} and \eqref{eqn: rate maximization special 22} holds with equality and the data rate of the secondary user is zero. 
For the case when $P\eta > \max\{C_p\sigma_p^2 \gamma_p,C_s\sigma_s^2 \gamma_p\}$, we can verify that $x_{\min} < \tilde x^* < x_{\max}$ and we have the following lemma.

\begin{lem} \label{lem: constraints active}
    Suppose that problem \eqref{p: rate maximization special 2} is feasible and the optimal pinching-antenna position satisfies $x_{\min} < \tilde x^* < x_{\max}$, both constraints \eqref{eqn: rate maximization special 21} and \eqref{eqn: rate maximization special 22} hold with equalities with the optimal solution.
\end{lem}

\textit{Proof:} See Appendix \ref{appd: constraints active}. \hfill $\blacksquare$

Based on Lemmas \ref{lem: feasibility} and \ref{lem: constraints active}, we can derive the optimal solution of problem \eqref{p: rate maximization special 2}, as stated in the following proposition.

\begin{lem}
    Suppose that $P\eta > \max\{C_p\sigma_p^2 \gamma_p,C_s\sigma_s^2 \gamma_p\}$, the optimal power allocation coefficient $\alpha_p$ that maximizes the objective function value and meanwhile guaranteeing the constraints are satisfied is given by
    \begin{align} \label{eqn: optimal alpha1}
        \alpha_p^* = \max\left\{ \frac{P\eta\gamma_p + C_p \sigma_p^2 \gamma_p}{P\eta(1 + \gamma_p)}, \frac{P\eta\gamma_p + C_s \sigma_s^2 \gamma_p}{P\eta (1 + \gamma_p)} \right\}.
    \end{align}
\end{lem}

\textit{Proof:} First, according to Lemma \ref{lem: constraints active}, with the optimal solutions, $\tilde x$ and $\alpha_p$ satisfy 
\begin{align} 
    (\tilde x - x_p)^2 &= \frac{P\eta\alpha_p (1+\gamma_p) - P\eta\gamma_p - C_p\sigma_p^2 \gamma_p}{\sigma_p^2\gamma_p}.\label{eqn: relationship of optimal solutions1} \\
    (\tilde x - x_s)^2 &= \frac{P\eta\alpha_p (1+ \gamma_p)  - P\eta\gamma_p - C_s\sigma_s^2 \gamma_p}{\sigma_s^2\gamma_p}.\label{eqn: relationship of optimal solutions2}
\end{align}
Next, by substituting \eqref{eqn: relationship of optimal solutions1} and \eqref{eqn: relationship of optimal solutions2} into the objective function \eqref{eqn: rate sa obj} and based on Lemma \ref{lem: constraints active}, problem \eqref{p: rate maximization special 2} can be equivalently simplified as 
\begin{subequations} \label{p: simplified problem with alpha1}
    \begin{align} 
        \underset{\alpha_p}{\rm{maximize}} ~ &\frac{\gamma_p(1-\alpha_p)}{(1 + \gamma_p) \alpha_p  -\gamma_p}\\
        {\rm{subject~to}} ~& B \leq \alpha_p \leq 1,
    \end{align}
\end{subequations}
where $$B \triangleq \max\left\{ \frac{P\eta\gamma_p \!+\! C_p \sigma_p^2 \gamma_p}{P\eta(1 + \gamma_p)}, \frac{P\eta\gamma_p \!+\! C_s \sigma_s^2 \gamma_p}{P\eta (1 + \gamma_p)} \right\}.$$
Let's define the objective function as 
\begin{align}
    f(\alpha_p) \triangleq \frac{\gamma_p(1-\alpha_p)}{(1 + \gamma_p) \alpha_p  -\gamma_p}.
\end{align}
The first-order derivative of $f(\alpha_p)$ is given by
\begin{align}
    f'(\alpha_p) = \frac{-\gamma_p}{\big((1 + \gamma_p) \alpha_p  - \gamma_p\big)^2}.
\end{align}
As seen, $f'(\alpha_p) < 0$ for $\alpha_p$ in the feasible set of problem \eqref{p: simplified problem with alpha1}. Therefore, the optimal $\alpha_p$ is given by \eqref{eqn: optimal alpha1}.
This completes the proof.
\hfill $\blacksquare$

With the optimal power allocation coefficient $\alpha_p^*$, the optimal pinching-antenna position $\tilde x$ can be determined based on Lemma \ref{lem: constraints active}, \eqref{eqn: relationship of optimal solutions1} and \eqref{eqn: relationship of optimal solutions2}. 
In particular, when $x_p \leq x_s$, the optimal pinching-antenna position $\tilde x^*$ is given by
\begin{align} \label{eqn: optimal pinching antenna location 1}
    \tilde x^* =  \beta_p + x_p, ~\textrm{or},~ \tilde x^* = x_s - \beta_s,
\end{align}
where $\beta_p = \big(\big|\frac{P\eta\alpha_p^* (1+\gamma_p) - P\eta\gamma_p - C_p\sigma_p^2 \gamma_p}{\sigma_p^2\gamma_p}\big|\big)^{\frac{1}{2}}$ and $\beta_s = \big(\big|\frac{P\eta\alpha_p^* ( 1+\gamma_p)  - P\eta\gamma_p - C_s\sigma_s^2 \gamma_p}{\sigma_s^2\gamma_p}\big|\big)^{\frac{1}{2}}$. 
When $x_p \geq x_s$, the optimal pinching-antenna position $\tilde x^*$ is given by
\begin{align} \label{eqn: optimal pinching antenna location 2}
    \tilde x^* = x_p - \beta_p, ~\textrm{or},~ \tilde x^* = \beta_s + x_s.
\end{align}

Summarizing the above analyses, the optimal solutions of problem \eqref{p: rate maximization special 2} can be obtained from the following three cases:
\begin{enumerate}
    \item When $P\eta = C_p\sigma_p^2 \gamma_p \geq C_s\sigma_s^2 \gamma_p$, the optimal pinching-antenna position is $\tilde x^* = x_p$, and the optimal power allocation coefficients are $\alpha_p^* = 1$ and $\alpha_s^* = 0$.
    \item When $P\eta = C_s\sigma_s^2 \gamma_p \geq C_p\sigma_p^2 \gamma_p$, the optimal pinching-antenna position is $\tilde x^* = x_s$, and the optimal power allocation coefficients are $\alpha_p^* = 1$ and $\alpha_s^* = 0$.
    \item When $P\eta > \max\{C_p\sigma_p^2 \gamma_p,C_s\sigma_s^2 \gamma_p\}$, the optimal pinching-antenna position is given by \eqref{eqn: optimal pinching antenna location 2}, the optimal $\alpha_p^*$ is given by \eqref{eqn: optimal alpha1} and $\alpha_s^* = 1 - \alpha_p^* \in (0,1)$.
\end{enumerate}

\begin{rem}
    Consider a special case where the received noise powers at both users are identical, i.e., i.e., $\sigma_p^2 = \sigma_s^2$, and the two users have the same distance to the waveguide, implying, $C_p = C_s$. Under these conditions, it follows that $\beta_p = \beta_s$. Substituting these equalities into \eqref{eqn: optimal pinching antenna location 1} or \eqref{eqn: optimal pinching antenna location 2} yields:
    \begin{align}
        \tilde x^* =  \frac{x_p + x_s}{2}.
    \end{align}
    This result indicates that the optimal pinching-antenna position is precisely positioned at the midpoint between the two users on the waveguide.
\end{rem}

\begin{rem}
    To examine the impact of the distances between the users and the waveguide, i.e., $\sqrt{C_p}$ and $\sqrt{C_s}$, on the optimal pinching-antenna position, we consider the scenario where $\sigma_p = \sigma_s$. From \eqref{eqn: optimal pinching antenna location 1} and \eqref{eqn: optimal pinching antenna location 2}, we observe that when $C_p \geq C_s$, it follows that $\beta_p \leq \beta_s$. In this case, the pinching antenna will be placed closer to the primary user. Conversely, when $C_s \geq C_p$, we have $\beta_s \leq \beta_p$, and the pinching antenna will be placed closer to the secondary user. This suggests that the optimal antenna placement is closer to the user who is positioned further from the waveguide, in order to compensate the larger path loss due to the user's increased distance.
\end{rem}

\section{Simulation Results} \label{sec: simulation}
In this section, we evaluate the performance of the pinching-antenna systems and the proposed algorithms via computer simulations. The simulation parameters are chosen as follows. The noise power is set to $-70$ dBm, $d = 3$ m, $\Delta = \frac{\lambda}{2}$, and $n_{\rm neff} = 1.4$. The choices for the working frequency $f_c$, the total transmit power $P$, the QoS requirement of primary user (i.e., $\gamma_p$), the side length ($D$) of the square area, and the number of pinching antennas $N$ are specified in each figure. In the simulations, the results are obtained by averaging $20$ random user deployments. The stopping condition for the all the iteration-based schemes in this work is that the absolute value of the difference in the objective function between two successive iterations is less than $10^{-3}$.

\subsection{Performance Gain of Pinching-Antenna Systems Compared with the Conventional Fixed-Position Antenna Systems}
In this subsection, we validate the performance gain of the pinching-antenna system by comparing it with the conventional antenna system, where all the $N$ BS antennas have fixed positions. We first summarize the downlink NOMA transmission in the conventional antenna system.
Without loss of generality, we assume that the BS antennas are deployed right above the centroid of the service area with a height $d$. The position of the $n$-th antenna is denoted by $\ensuremath{\bm \bar \psi}_n = [\bar x_n,0,d], \forall n \in \Nset$. The spacing between two neighboring antennas is set as $\Delta = \frac{\lambda}{2}$ to avoid antenna coupling. According to the spherical wave channel model \cite{zhang2022beam}, the channel vector between the fixed antennas and user $m$ is given by
\begin{align} \label{eqn: h convention}
    \hb_m^{\rm Conv} = \bigg[ \frac{\eta^{\frac{1}{2}} e^{-j \frac{2\pi}{\lambda}\|\bm{\psi}_m - \boldsymbol{\bar\psi}_1\|}}{\|\boldsymbol{\psi}_m - \boldsymbol{\bar\psi}_1\|},..., \frac{\eta^{\frac{1}{2}} e^{-j \frac{2\pi}{\lambda}\|\boldsymbol{\psi}_m - \boldsymbol{\bar\psi}_N\|}}{\|\boldsymbol{\psi}_m - \boldsymbol{\bar\psi}_N\|}\bigg]^\top\!.
\end{align}
Note that unlike the channel vector in \eqref{eqn: h pinching}, which is adjustable by changing the positions of the pinching antennas, $\hb_m^{\rm Conv}$ in \eqref{eqn: h convention} is fixed. 
Let $\vb_p \in \Cs^{N}$ and $\vb_s \in \Cs^{N}$ denote the associated beamformers for the primary user and the secondary user, respectively. Then, the QoS-aware problem can be formulated as 
\begin{subequations} \label{p: qos fixed antenna}
    \begin{align}
        \underset{\vb_p, \vb_s}{\rm{maximize}}~& \frac{\|\hb_s^\top \vb_s\|^2}{\sigma_s^2}\\
        \st ~&\frac{\|\hb_p^\top \vb_p\|^2}{\|\hb_p^\top \vb_s\|^2 + \sigma_p^2} \geq \gamma_p,\\
        &\frac{\|\hb_s^\top \vb_p\|^2}{\|\hb_s^\top \vb_s\|^2 + \sigma_s^2} \geq \gamma_p,\\
        &\|\vb_p\|^2 + \|\vb_s\|^2 \leq P,
    \end{align}
\end{subequations}
which can be optimally solved by using semidefinite relaxation approach \cite{xu2017joint}. 

\begin{figure}[!t]
	\centering
	\includegraphics[width=0.86\linewidth]{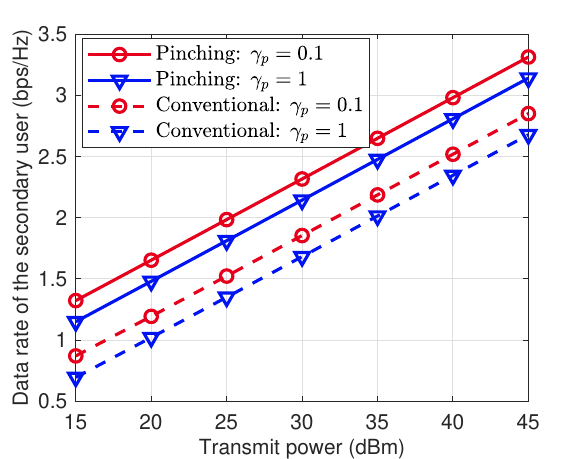}\\
        \captionsetup{justification=justified, singlelinecheck=false, font=small}	
        \caption{The achievable data rate of the secondary user versus the total transmit power $P$ with the number of pinching antennas $N = 2$ and $f_c = 28$ GHz.} \label{fig: rate conv power} 
\end{figure} 

\begin{figure}[!t]
	\centering
	\includegraphics[width=0.86\linewidth]{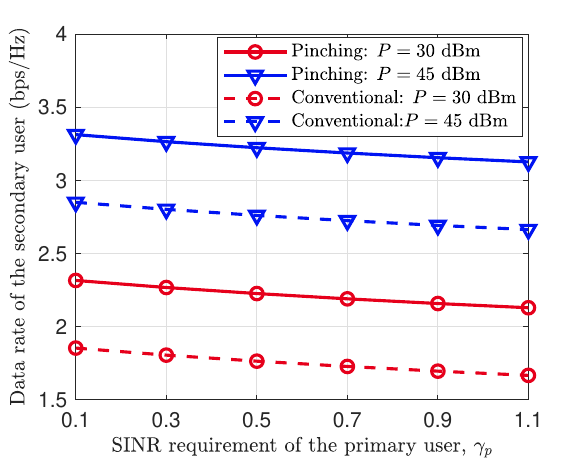}\\
        \captionsetup{justification=justified, singlelinecheck=false, font=small}	
        \caption{The achievable data rate of the secondary user versus the SINR requirement of the primary user with the number of pinching antennas $N = 2$ and $f_c = 28$ GHz.} \label{fig: rate conv gamma} 
\end{figure} 

Fig. \ref{fig: rate conv power} is provided to numerically demonstrate that the pinching-antenna system consistently achieves a higher data rate for the secondary user compared to the conventional fixed-position antenna system across all adopted total transmit power levels. This superior performance is primarily attributed to the pinching antenna system’s ability to dynamically adjust antenna positions, ensuring that the pinching antennas can be flexibly deployed to the ideal positions which reduce the large-scale path loss. Furthermore, as the transmit power increases, the advantages of the pinching-antenna system introduced by the flexibility in adjusting the antenna positions also brings a widening performance gap between the two systems. In Fig. \ref{fig: rate conv gamma}, the performance gain of the pinching-antenna system is further confirmed by comparing the ergodic achievable data rate of the secondary user in the pinching-antenna system and the conventional antenna system versus the SINR requirement of the primary user.
These findings underscore the potential of pinching antenna systems in improving user performance and enhancing the overall network efficiency, particularly in scenarios demanding high data throughput and robust interference mitigation.

\subsection{Performance of the Proposed BCD-SCA Algorithm}
\begin{figure}[t]
	\centering
	\subfigure[$f_c = 6$ GHz]
	{\includegraphics[width=0.86\linewidth]{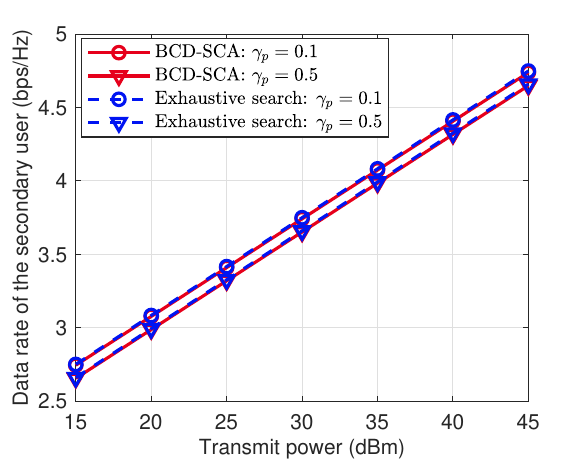}\label{fig: f6}}\\
	\subfigure[$f_c = 28$ GHz]{
		\includegraphics[width=0.86\linewidth]{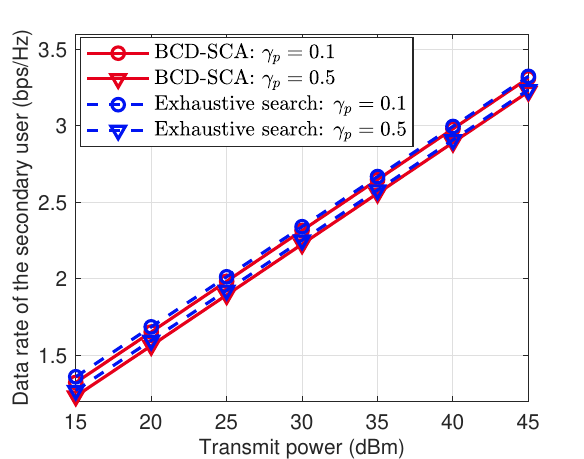}\label{fig: f28}}
        \captionsetup{justification=justified, singlelinecheck=false, font=small}	
	\caption{\small The achievable data rate of the secondary user versus the total transmit power $P$ with $\gamma_p = 0.5$, $D = 5$ m, and $N = 2$.} 
	\label{fig: bcd-exhaustive}
\end{figure}

\begin{figure}[t]
	\centering
	\subfigure[$N = 2$ ]
	{\includegraphics[width=0.86\linewidth]{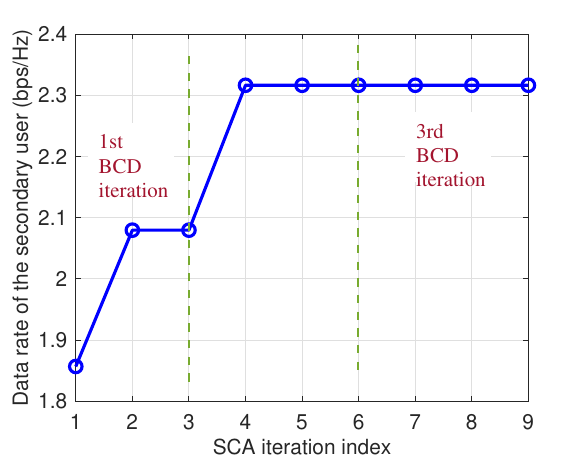}\label{fig: low snr}}\\
	\subfigure[$N = 10$]{
		\includegraphics[width=0.86\linewidth]{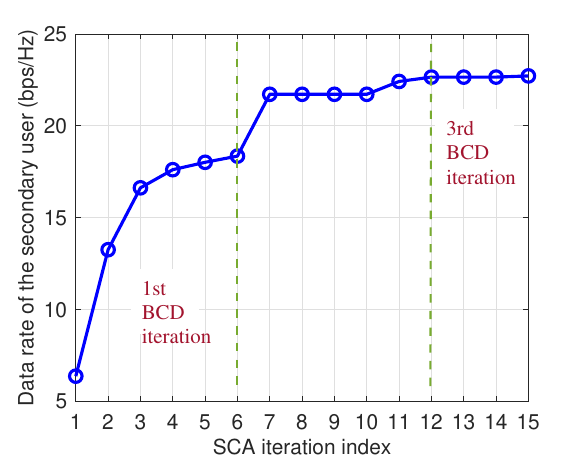}\label{fig: high snr}}
        \captionsetup{justification=justified, singlelinecheck=false, font=small}	
	\caption{\small The convergence performance of the proposed BCD-SCA algorithm with $\gamma_p = 0.1$, $D = 5$ m, $f_c = 28$ GHz, and $P = 30$ dBm.}
	\label{fig : BCD-SCA convergence}
\end{figure}

To evaluate the effectiveness of the proposed BCD-SCA algorithm, we compare it against the exhaustive search method in Fig. \ref{fig: bcd-exhaustive}, which serves as a performance benchmark. Recall from Section \ref{sec: update of power allocation} that the global optimal solution to problem \eqref{p: rate maximization origin} can be obtained if the pinching-antenna positions are determined via an exhaustive search over the waveguide. Motivated by this observation, the exhaustive search method is conducted by discretizing the waveguide into steps of size $\frac{\lambda}{50}$ and evaluating all possible antenna positions at each step. For each of the position, the optimal power allocation coefficients are found according to equation \eqref{eqn: power allocation}, thereby guaranteeing the global optimal solution. 
Since the complexity of the exhaustive search approach grows rapidly with the number of pinching antennas, Fig. \ref{fig: bcd-exhaustive} focuses on the special case with $N=2$ pinching antennas, where an exhaustive search is still computationally tractable. The results show that the proposed BCD-SCA algorithm yields the data rate performance very close to that of the exhaustive search-based optimal solution. 
Overall, these results confirm the efficacy of the proposed BCD-SCA algorithm and validate the pinching-antenna position initialization scheme. Despite its much lower computational complexity, the BCD-SCA algorithm achieves near-optimal performance, making it a promising candidate for practical large-scale deployments where an exhaustive search would be prohibitively expensive.

{
\begin{table}[t] 
	\caption{Execution time (in seconds) of the considered schemes with $f_c = 28$ GHz, $\gamma_p = 0.1$ and $P = 40$ dBm.} 
	\label{tab: Execution time}  \renewcommand\arraystretch{1.5}\setlength{\tabcolsep}{1.53mm}{
	\begin{tabular}{
    >{\centering\arraybackslash}p{1.25cm}
    >{\centering\arraybackslash}p{1cm}
    >{\centering\arraybackslash}p{0.82cm}
    >{\centering\arraybackslash}p{0.82cm}
    >{\centering\arraybackslash}p{0.82cm}
    >{\centering\arraybackslash}p{1cm}
    >{\centering\arraybackslash}p{1cm}
    } 
		\hline 
		 & \multicolumn{6}{c}{\cellcolor{green!12} Side Length $D = 5$ m} \\  
		\cline{2-7}
		& $N=2$ & $N=4$ & $N=6$ & $N=8$ & $N=10$ & $N=12$\\  
		\hline
		BCD-SCA &$0.79$  &  $1.46$  &  $2.03$  &  $2.26$  &  $2.60$  &  $2.66$ \\
            \hline
		Exhaustive search & \multirow{2}{*}{$451.68$} & \multirow{2}{*}{-} & \multirow{2}{*}{-} & \multirow{2}{*}{-} & \multirow{2}{*}{-}  & \multirow{2}{*}{-}\\
		\hline 
				 & \multicolumn{6}{c}{\cellcolor{green!12}  Side Length $D = 10$ m} \\  
		\cline{2-7}
		& $N=2$ & $N=4$ & $N=6$ & $N=8$ & $N=10$ & $N=12$\\   
		\hline
		BCD-SCA &$1.51$  &  $3.19$  &  $3.99$  &  $4.52$  &  $4.90$  &  $4.98$\\
            \hline
		  Exhaustive search& \multirow{2}{*}{$1792.18$} & \multirow{2}{*}{-} & \multirow{2}{*}{-} & \multirow{2}{*}{-} & \multirow{2}{*}{-}  & \multirow{2}{*}{-}\\
		\hline 
	\end{tabular}} 
\end{table} }

To further validate the efficiency of the proposed BCD-SCA algorithm in reducing the computational complexity, we investigate its convergence behavior in Fig. \ref{fig : BCD-SCA convergence} under two configurations of number of pinching antennas, i.e., $N = 2$ (in  Fig. \ref{fig: low snr}) and $N = 10$ (in  Fig. \ref{fig: high snr}). For  better illustration, the evolution of the objective values over each iteration of the SCA-based sub-algorithm  (for updating the pinching-antenna positions) is plotted. Results show that the proposed algorithm converges quickly. For instance, when $N = 10$, only three outer iterations of the BCD-SCA algorithm are required to achieve convergence, and, within each iteration, the SCA-based sub-algorithm requires only six inner iterations to optimize the pinching-antenna positions. 
We further compare the execution times of the proposed BCD-SCA algorithm and the exhaustive search-based method under varying numbers of pinching antennas and communication area side lengths, as summarized in Table~\ref{tab: Execution time}. Due to the rapidly increasing complexity of the exhaustive search method with respect to the number of antennas, we limit its evaluation to the case of $N = 2$. All simulations are conducted on a workstation equipped with an Intel(R) Core(TM) i9-10900X CPU @ 3.70 GHz. As shown in Table~\ref{tab: Execution time}, although the execution time of the BCD-SCA algorithm increases with the number of antennas, it remains significantly lower than that of the exhaustive search-based method, while achieving comparable performance as illustrated in Fig.~\ref{fig: bcd-exhaustive}.

\begin{figure}[!t]
	\centering
	\includegraphics[width=0.86\linewidth]{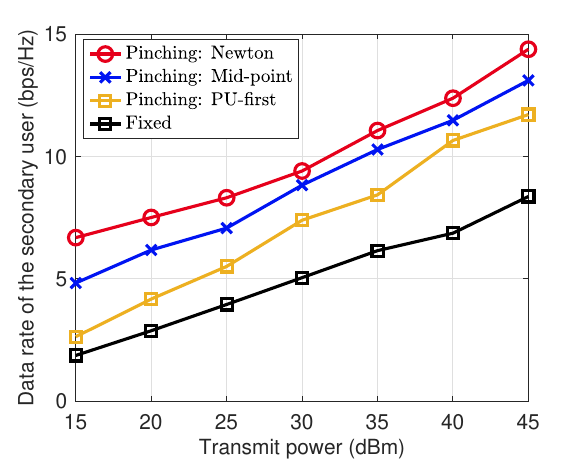}\\
        \captionsetup{justification=justified, singlelinecheck=false, font=small}	
        \caption{The achievable data rate of the secondary user versus the total transmit power $P$ under different initialization schemes, with the number of pinching antennas $N = 4$, $\gamma_p = 0.1$, and $f_c = 28$ GHz.} \label{fig: rate initialization} 
\end{figure} 
To evaluate the impact of initialization strategies on algorithm performance, we compare four different schemes: Newton-based, Mid-point, PU-first, and Fixed. In the Mid-point scheme, the first pinching antenna is placed at the midpoint between the two users on the waveguide, i.e., $\tilde{x}_1 = (x_p + x_s)/2$, with the remaining antennas sequentially positioned at fixed intervals of $\Delta$. In the PU-first scheme, the first antenna is initialized at the position closest to the primary user's location ($\tilde{x}_1 = x_p$) on the waveguide, and the others follow similarly. The Fixed scheme uses pre-defined static positions, independent of user locations. Simulation results, as shown in Fig. \ref{fig: rate initialization}, demonstrate that the Newton-based initialization consistently yields higher data rates across all transmission power levels. This performance gain is attributed to the fact that the Newton scheme strategically places the first pinching antenna by approximately minimizing the free-space path loss to the user positions. As a result, it provides a more favorable starting point for subsequent antenna refinement. These observations highlight the critical role of  adaptive initialization in achieving improved performance in the proposed BCD-SCA algorithm.

\subsection{Performance Comparison Between NOMA and OMA Transmission Schemes in Pinching-Antenna Systems}
\begin{figure}[t]
	\centering
	\subfigure[$f_c = 6$ GHz]
	{\includegraphics[width=0.9\linewidth]{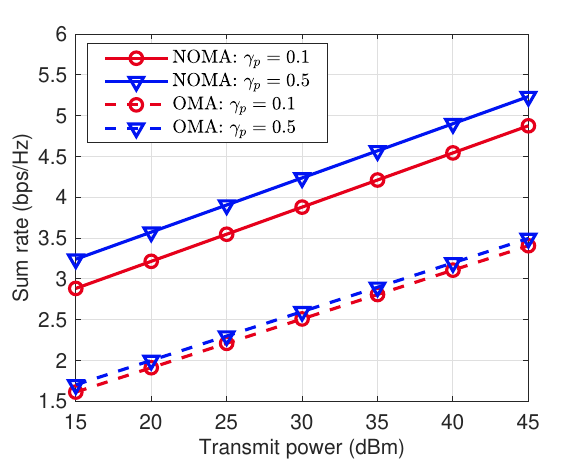}\label{fig: sum d5}}\\
	\subfigure[$f_c = 28$ GHz]{
		\includegraphics[width=0.9\linewidth]{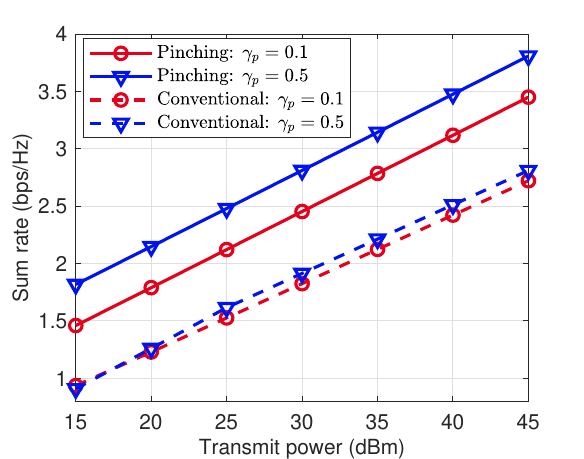}\label{fig: sum d10}}
        \captionsetup{justification=justified, singlelinecheck=false, font=small}	
	\caption{\small Sum rate comparison of the proposed NOMA scheme and OMA scheme versus the total transmit power with $N = 2$, and $D = 5$ m.} 
	\label{fig: noma oma}
\end{figure}

Next, we evaluate the performance gains of the proposed NOMA transmission scheme relative to the OMA transmission scheme from Section \ref{sec: OMA}. Under the OMA scheme, the two users are served individually, and the corresponding rate-maximization problem for each user follows the formulation in \cite{xu2024rate}, where the proposed algorithm in \cite{xu2024rate} can be applied. To quantify the differences, Fig. \ref{fig: noma oma} depicts the sum rates for the NOMA and OMA schemes versus the total transmit power. As illustrated, the NOMA transmission scheme consistently outperforms its OMA counterpart. The primary reason is that NOMA exploits additional degree of freedom by power-domain multiplexing to serve both users simultaneously, whereas OMA orthogonally allocates resources to each user, limiting its efficiency. Consequently, the additional degree of freedom in NOMA substantially improves the sum rate compared to the OMA approach.

\begin{figure}[!t]
	\centering
	\includegraphics[width=0.9\linewidth]{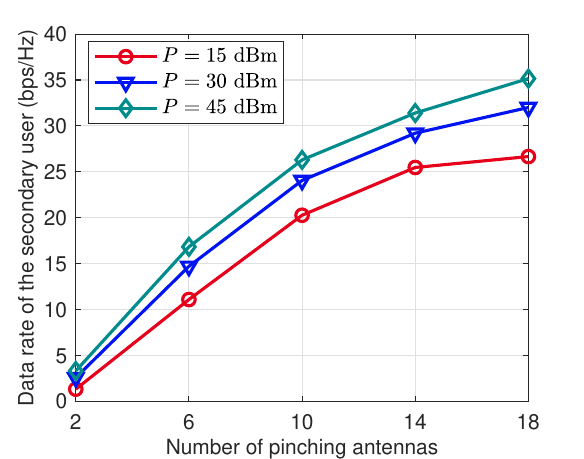}\\
        \captionsetup{justification=justified, singlelinecheck=false, font=small}	
        \caption{The achievable data rate of the secondary user versus different number of pinching antennas with $\gamma_p = 0.1$, $f_c = 28$ GHz, and $D = 5$ m.} \label{fig: rate gamma}
\end{figure}

\subsection{Impact of System Parameters on the Performance of Pinching-Antenna Systems}

We further investigate how the number of pinching antennas influences the performance of the NOMA transmission scheme. Fig. \ref{fig: rate gamma} illustrates the achievable data rate of the secondary user versus the number of pinching antennas, denoted by $N$. As observed, the data rate for the secondary user increases monotonically with $N$. This trend is due to the fact that each additional pinching antenna enriches the channel diversity and enhances the effective gain perceived at the secondary user, thereby providing higher signal strength and better interference management. Consequently, the spectral efficiency improves with more pinching antennas deployed along the waveguide, leading to higher achievable data rate. This result highlights the importance of optimizing the number of pinching antennas in practical systems to strike a balance between performance gains and hardware/implementation constraints. Besides, it is worthy to point out that adding or removing antennas in conventional antenna systems is not
straightforward, whereas the pinching-antenna system offers superior flexibility to reconfigure the antenna system.

\begin{figure}[!t]
	\centering
	\includegraphics[width=0.9\linewidth]{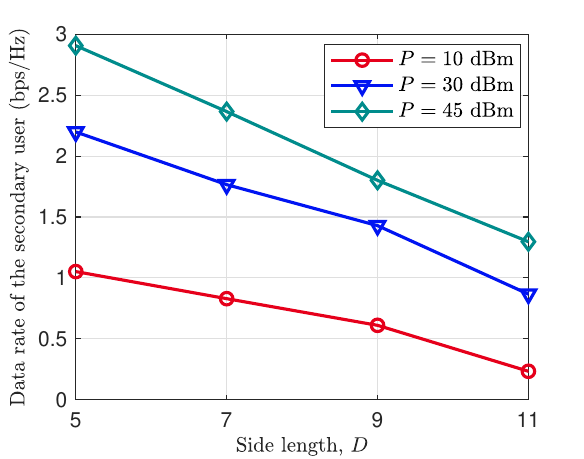}\\
        \captionsetup{justification=justified, singlelinecheck=false, font=small}	
        \caption{The achievable data rate of the secondary user versus different side length with $N = 2$, and $f_c = 28$ GHz}. \label{fig: rate side length} 
\end{figure}

Fig. \ref{fig: rate side length} shows the impact of the side length $D$ (i.e., the dimension of the user deployment region) on the achievable data rate of the secondary user under the proposed NOMA scheme. Our simulation results indicate that as $D$ becomes larger, the data rate of the secondary user decreases. This phenomenon can be understood from a geometric perspective: with a larger deployment area, both users can potentially be located further apart, and the pinching antennas may need to span longer distances to reach the secondary user with sufficient signal strength. Consequently, the path loss experienced by the secondary user is likely to increase, which in turn diminishes the effective channel gain and reduces the secondary user’s data rate.
Moreover, larger $D$ expands the search space for pinching-antenna placement, making it increasingly important to optimize their positions so as to counteract the enhanced path loss and maintain a high spectral efficiency. In these scenarios, a robust pinching-antenna position optimization strategy becomes critical, as it can strategically position the antennas to compensate for the larger distances and maintain strong channel conditions for the secondary user. This underscores that the effectiveness of the pinching-antenna positioning scheme plays a central role in enhancing the achievable data rate in large network coverage settings.

\begin{figure}[!t]
	\centering
	\includegraphics[width=0.9\linewidth]{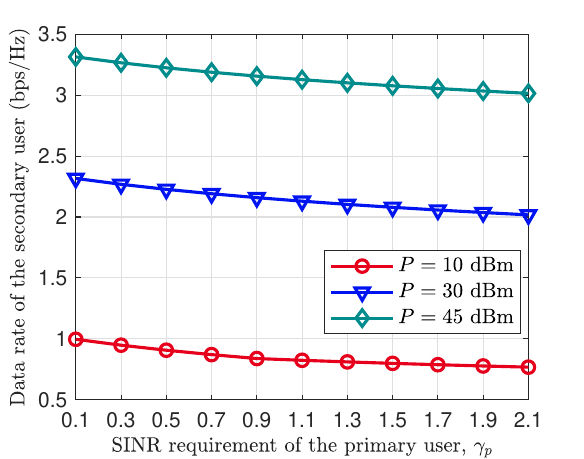}\\
        \captionsetup{justification=justified, singlelinecheck=false, font=small}	
        \caption{The achievable data rate of the secondary user versus different $\SINR$ requirement of the primary user with $N = 2$,  $f_c = 28$ GHz, and $D = 5$ m.} \label{fig: rate gamma12} 
\end{figure} 

We further evaluate how the $\SINR$ requirement, i.e., $\gamma_p$, of primary user, influences the achievable data rate of the secondary user. As depicted in Fig. \ref{fig: rate gamma12}, increasing $\gamma_p$ reduces the data rate of the secondary user. This decline is a natural consequence of the larger $\SINR$ requirement of the primary user, which in turn compels the system to allocate more resources (e.g., transmit power or more favorable antenna configurations) toward satisfying the primary user’s QoS requirements. Since the total transmit power is finite, the secondary user inevitably experiences reduced received power. Consequently, the synergy between primary user’s high QoS requirements and finite system resources explains why the secondary user’s data rate reduces as $\gamma_p$ increases. These results underscore the trade-off in multi-user systems between ensuring high reliability for one user versus sustaining higher data rate for other users.

\begin{figure}[!t]
	\centering
	\includegraphics[width=0.86\linewidth]{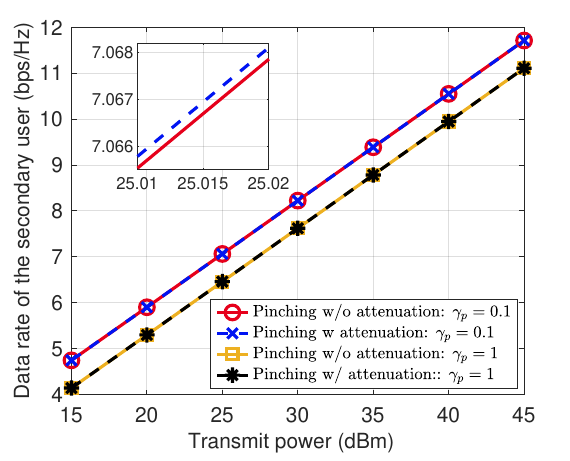}\\
        \captionsetup{justification=justified, singlelinecheck=false, font=small}	
        \caption{The achievable data rate of the secondary user versus the total transmit power $P$ under schemes with and without in-waveguide attenuation, with the number of pinching antennas $N = 4$, $D = 10$ m, and  $f_c = 28$ GHz.} \label{fig: rate attenuation} 
\end{figure}
To assess the practical impact of in-waveguide attenuation on system design, we compare two schemes: one that incorporates in-waveguide attenuation in both the optimization and rate evaluation stages, and another that ignores it only during the optimization stage but still accounts for it when computing the final achievable data rate. Both schemes are solved using the proposed BCD-SCA algorithm. In the simulation, the in-waveguide attenuation is set as $\alpha = 0.08$ dB/m, and the results are shown in Fig. \ref{fig: rate attenuation}. As observed, the performance gap between the two schemes is negligible across all considered transmit power levels. This demonstrates that ignoring in-waveguide attenuation during the design phase introduces minimal performance loss, especially when the communication area is reasonably bounded. These findings are consistent with the theoretical insights reported in \cite{xu2025pinching}, and they support the use of a simplified channel model to facilitate analytical derivations.

\begin{figure}[!t]
	\centering
	\includegraphics[width=0.8\linewidth]{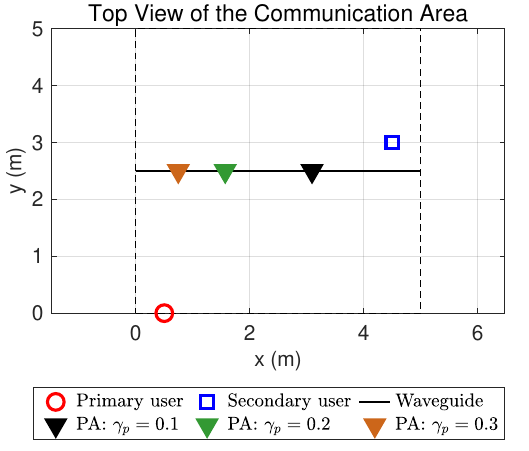}\\
        \captionsetup{justification=justified, singlelinecheck=false, font=small}	
        \caption{Optimal pinching antenna (PA) placement versus different target SINR requirement of the primary user with $N = 1$,  $f_c = 28$ GHz, and $D = 5$ m.} \label{fig: rate gamma1} 
\end{figure} 

To provide further insight of the impact of user distribution and QoS requirements on system design, we present a top-view visualization of a single pinching-antenna system in Fig. \ref{fig: rate gamma1}. In this simulation, we fix the positions of the primary and secondary users at $[0.5,0,0]$ and $[4.5,3,0]$, respectively. The optimal pinching antenna location is obtained by using the closed-form expression derived in \eqref{eqn: optimal pinching antenna location 2}. We then compare the resulting antenna positions for three different SINR requirements of the primary user, namely, $\gamma_p = 0.1$, $0.2$, and $0.3$.
As observed in Fig. \ref{fig: rate gamma1}, when $\gamma_p$ is small, the optimal antenna is placed closer to the secondary user to improve its data rate. As $\gamma_p$ increases, the antenna gradually shifts toward the primary user. This repositioning helps meet the primary user's more stringent SINR requirement while ensuring successful SIC at the secondary user at a price of reduced secondary user data rate. These trends clearly demonstrate how spatial placement of the antenna serves as a flexible design degree of freedom for balancing the performance requirements of different users.

\section{Conclusion} \label{sec: conclusion}
In this paper, we have investigated a downlink pinching-antenna system serving two users using the NOMA scheme. The system leverages the flexibility of pinching antennas deployed along a waveguide to dynamically optimize the pinching-antenna positions and the power allocation coefficients, offering advantages over conventional fixed-position systems. We addressed the joint power allocation and pinching-antenna position optimization problem by decoupling it into two subproblems and solving them using a BCD-SCA algorithm. Furthermore, we have explored the case with a single pinching antenna serving two users using NOMA. By properly exploiting the structure of the associated problem, we have derived the optimal solution in closed-form. Numerical simulations validated the performance of the proposed NOMA scheme and the BCD-SCA algorithm. The results showed that the pinching-antenna system outperforms traditional systems, demonstrating its potential as a flexible solution for next-generation wireless communication systems. This work highlights the promise of pinching-antenna systems in adapting to dynamic environments and optimizing performance in next-generation wireless networks. 

There are several important directions for future research. To further enhance the system's spectral efficiency, it is crucial to develop more flexible power allocation strategies across multiple pinching antennas.
Furthermore, extending the NOMA-assisted pinching-antenna system design to a multi-waveguide scenario is a promising direction, as it enables higher spatial diversity. However, it also introduces challenges such as inter-waveguide interference and more complex joint optimization, which deserve further investigation.
In addition, extending the current two-user NOMA framework to more general multi-user scenarios is another important and promising direction. While such an extension can further improve spectrum utilization by serving more users simultaneously, it also significantly increases the dimensionality of the optimization space and introduces more intricate interference and SIC constraints. Tackling these challenges will be essential for fully unleashing the potential of NOMA-assisted pinching-antenna systems.

\begin{appendices}
    \section{Proof of Lemma \ref{lem: location of x}} \label{appd: lemma location of x}
    We prove Lemma \ref{lem: location of x} by contradiction. First, assume that the optimal value $\tilde{x}^* < x_{\min}$. In this case, we can increase $\tilde{x}^*$ such that $\tilde{x}^* = x_{\min}$. It can be verified that, with $\tilde{x}^* = x_{\min}$, constraints \eqref{eqn: rate maximization special 21} and \eqref{eqn: rate maximization special 21} are still satisfied, while the value of the objective function increases. 
    Similarly, if $\tilde{x}^* > x_{\max}$, we can set $\tilde{x}^* = x_{\max}$, ensuring that the constraints \eqref{eqn: rate maximization special 21} and \eqref{eqn: rate maximization special 21} still hold, and the value of the objective function also increases.
    In conclusion, the optimal position of the pinching antenna must satisfy $x_{\min} \leq \tilde{x}^* \leq x_{\max}$. This completes the proof.   \hfill $\blacksquare$

    \section{Proof of Lemma \ref{lem: constraints active}} \label{appd: constraints active}
    Lemma \ref{lem: constraints active} can be proved by contradiction. First of all, it can be verified that at least one of the two constraints \eqref{eqn: rate maximization special 21} and \eqref{eqn: rate maximization special 22} holds with equality with the optimal solution. 
    Otherwise, one can directly decrease $\alpha_p$ until one of them holds with equality, while the objective function value decreases.
    In the following, we will prove that both constraints \eqref{eqn: rate maximization special 21} and \eqref{eqn: rate maximization special 22} should hold with equality with the optimal solution. To this end, we consider the following two cases:
    \begin{itemize}
        \item[{Case I:}] Constraint \eqref{eqn: rate maximization special 21} holds with inequality while constraint \eqref{eqn: rate maximization special 22} holds with equality. In this case, one can change the value of $\tilde x$, such that $(\tilde x - x_p)^2$ increases and the value of the left hand side of constraint \eqref{eqn: rate maximization special 21} decreases. Then, according to Lemma \ref{lem: location of x}, the value of $(\tilde x - x_s)^2$ should decrease and the value of the left hand side of constraint \eqref{eqn: rate maximization special 22} increases. Consequently, both constraints \eqref{eqn: rate maximization special 21} and \eqref{eqn: rate maximization special 22} hold with inequality. Next, one can decrease the value of $\alpha_p$ such that one of the two constraints holds with equality, and meanwhile the value of the objective function increases. Then, one can repeat the above procedure by gradually changing the values of $x$ and $\alpha_p$ until both constraints hold with equality.
        
        \item[{Case II:}] Constraint \eqref{eqn: rate maximization special 21} holds with equality while constraint \eqref{eqn: rate maximization special 22} holds with inequality. In this case, similar approach as in Case I can be applied to show that both constraints should hold with equality with the optimal solution. 
    \end{itemize}
    In summary, as long as problem \eqref{p: rate maximization special 2} is feasible, both constraints \eqref{eqn: rate maximization special 21} and \eqref{eqn: rate maximization special 22} hold with equality with the optimal solution.   \hfill $\blacksquare$

\end{appendices}


\smaller[1]

\end{document}